\documentclass{ws-ijmpa}
\usepackage{cite} 
\usepackage{epsfig}
\usepackage{color}

\begin{document}

%
%

\title{Renormalisation running of masses and mixings in UED models}

\author{A.S. Cornell}
\address{National Institute for Theoretical Physics; School of Physics, \\
University of the Witwatersrand, Wits 2050, South Africa\\
alan.cornell@wits.ac.za}

\author{Aldo Deandrea}
\address{Universit\'e de Lyon, F-69622 Lyon, France; \\
Universit\'e Lyon 1, CNRS/IN2P3, UMR5822 IPNL, F-69622 Villeurbanne Cedex, France\\
deandrea@ipnl.in2p3.fr}

\author{Lu-Xin~Liu}
\address{National Institute for Theoretical Physics; School of Physics, \\
University of the Witwatersrand, Wits 2050, South Africa\\
luxin.liu9@gmail.com}

\author{Ahmad Tarhini}
\address{Universit\'e de Lyon, F-69622 Lyon, France; \\
Universit\'e Lyon 1, CNRS/IN2P3, UMR5822 IPNL, F-69622 Villeurbanne Cedex, France\\
tarhini@ipnl.in2p3.fr}

\maketitle

\begin{abstract}
We review the renormalisation group evolution of quark and lepton masses, mixing angles and phases both in the UED extension of the Standard Model and of the Minimal Supersymmetric Standard Model. We consider two typical scenarios:  all matter fields propagating in the bulk, and matter fields constrained to the brane. The resulting renormalisation group evolution equations in these scenarios are compared with the existing results in the literature, together with their implications.

\end{abstract}

\ccode{11.10.Hi, 11.10.Kk, 12.60.Jv, 14.60.Pq, 12.15.Ff}

%
%

\section{Introduction}\label{sec:1}	

\par The Standard Model (SM), by meeting all confrontations with experiments, stands as a remarkably simple parameterisation of known physics. Yet it has many unsatisfactory aspects which leads to a belief that there must exist a simpler underlying structure of which the SM is the low energy piece. This structure is believed to make its appearance at much higher energies, where we can approach this with  renormalisation group techniques to extrapolate the SM parameters to the unexplored scales \cite{RGEearly}.  

\par Recalling that in the SM, the runnings of the gauge, Yukawa and quartic scalar couplings are logarithmic with the energy scale, the gauge couplings do not all meet at a point, but do tend to unify near $10^{15}$ GeV. Extensions to the SM such as extra-dimensional scenarios accessible to SM fields have the virtue, thanks to the couplings now having a power law running, of bringing the unification scale down to an explorable range \cite{DDG, DDG1}. Note that many other extensions to the SM exist which alter the runnings in different ways, such as supersymmetry (SUSY), where a range of new particles ensure the gauge couplings do meet at a point, but runnings remain logarithmic \cite{Pati:1974yy, Georgi:1974sy, Fritzsch:1974nn, Fritzsch:1974nm}.

\par The story of extra-dimensional physics can be thought to begin in the 1920s with Kaluza (1919) and Klein (1926) \cite{Klein:1926tv} who had the idea to add a fifth dimension to unify the only two forces known at that time. Later in the 1970s and 1980s‚ the birth of supergravity and superstring theories renewed the interest in extra-dimensional models. However these dimensions are expected to be very small ($M_P \sim 10^{-35} m$) and will not be probed by experiments any time soon. However, beginning in the 1990s new extra-dimensional scenarios which could be larger than the Planck length appeared. Antoniadis \cite{Antoniadis:1990ew} proposed $TeV^{-1}$ scale extra-dimensions to explain SUSY breaking, and in order to solve the hierarchy problem the Large Extra Dimensions approach was introduced by Arkani-Hamed, Dimopoulos and Dvali (ADD) \cite{ArkaniHamed:1998rs, Antoniadis:1998ig, ArkaniHamed:1998nn, ArkaniHamed:1999dz}. In these models, the metric is flat and the strength of the gravitational interaction is diluted, which leads to interesting consequences for low-energy phenomenology.

\par Another approach was introduced by Randall and Sundrum \cite{Randall:1999ee} with only one curved extra-dimension (warped extra-dimensions). The new dimension is compactified on a finite interval $0 \le y \le L$, with the endpoints of the interval being 3-branes. The metric of this space is not flat, where the gravity fields propagating in the fifth dimension suffer exponential suppression and live on a different brane to the SM particles.

\par Extra-dimensional models lead to many phenomenological implications which can be tested at colliders and also can be used as a tool to answer many issues in the SM, such as the hierarchy problem \cite{ArkaniHamed:1998rs, Antoniadis:1998ig, ArkaniHamed:1998nn, ArkaniHamed:1999dz, Randall:1999ee}; TeV scale extra-dimensional scenarios giving rise to new SUSY breaking mechanisms \cite{Antoniadis:1990ew}; the generation of neutrino mass and new sources of CP violation \cite{ArkaniHamed:1998vp, ArkaniHamed:1999dc, Lillie:2003sq}; the unification without SUSY with suppression of proton decay \cite{Dienes:1998vh, Dienes:1998vg, Carena:2003fx, Randall:2001gb}; triggering electroweak symmetry breaking without a Higgs boson \cite{Cacciapaglia:2004rb, Csaki:2003sh, Csaki:2003zu, Csaki:2003dt, Nomura:2003du, Davoudiasl:2004pw, Davoudiasl:2003me, Barbieri:2003pr}; providing cosmologically viable dark matter candidates \cite{Servant:2002aq, Cheng:2002iz, Appelquist:2000nn} and many other applications related to black holes and gravity \cite{Dimopoulos:2001hw, Giddings:2001bu}.

\par With the Large Hadron Collider (LHC) now up and running, exploration of the realm of new physics that may operate at the TeV scale has begun \cite{Antoniadis, Accomando:1999sj}. Among these models with extra spatial dimensions the Universal Extra-Dimension (UED) model makes for an interesting TeV scale physics scenario \cite{Antoniadis:1990ew, Delgado:1998qr}; as it features a tower of Kaluza-Klein (KK) states for each of the SM fields, all of which have full access to the extended space-time manifold \cite{Antoniadis, Appelquist:2000nn}. This particular scenario has recently been extensively studied in the literature \cite{Arkani-Hamed:2000, Appelquist:2001nn, Bhattacharyya:2006ym, Buras:2003mk, Buras:2002ej, Hooper:2007qk, Colangelo:2007jy, Datta:2010us} and has been a fruitful playground for addressing a variety of puzzles in the SM.

\par We therefore collect in a comprehensive manner and in one place the necessary tools for making renormalisation group analyses of the SM and Minimal Supersymmetric SM (MSSM) UED extensions. Note that we review only these particular UED scenarios here, as alternative warped or higher-dimensional UED models are largely unexplored in the literature, or require different toolsets, as shall be briefly discussed later. The observable parameters of the SM are: 6 quark masses, 3 lepton masses, 4 parameters of the Cabibbo-Kobayashi-Maskawa (CKM) matrix \cite{Kobayashi:1973fv, Cabibbo:1963yz} and 3 gauge couplings. The Renormalisation Group Equations (RGEs) for the CKM matrix being obtained from the RGEs for the Yukawa couplings. This can also be extended to include neutrino masses and mixings possible in the leptonic sector.

\par In this review we first introduce the various models and their varieties we shall consider (section \ref{sec:2} and section \ref{sec:3} for the supersymmetric extensions to this), next reviewing the RGEs for the gauge couplings constants (section \ref{sec:4}) and Yukawa couplings for the SM and UED scenarios and 5-dimensional Minimal Supersymmetric SMs (5D MSSM) (section \ref{sec:5}). This shall be followed by a review of the CKM parameters evolution (sections \ref{sec:7}). Extensions to massive neutrino scenarios and their mixings evolution will follow in section \ref{sec:9}. With a summary and the prospects for future research directions in section \ref{sec:10}.

%
%

\section{The UED Standard Model}\label{sec:2}

\par The UED model places particles of the SM in the bulk of one or more compactified extra dimensions \cite{Pomarol:1998sd}. In our case we have a single flat extra dimension of size $R$, compactified on an $S_1/Z_2$ orbifold. As such we will have an infinite tower of KK modes with the zero modes corresponding to the SM states. These KK modes are in the TeV scale and modify the running of the RGEs at relatively low energy scales \cite{Delgado:1999sv}. The UED model, like any higher dimensional theory, is an effective field theory which is valid up to some scale $\Lambda$, at which a new physics theory emerges. As a result, once the KK states are excited, these couplings exhibit power law dependencies on $\Lambda$. This can be illustrated if $\Lambda R \gg 1$, to a very good accuracy, the generic SM beta function is shown to have the power law evolution behaviour \cite{Bhattacharyya:2006ym}:
\begin{equation}
\beta^{4D} \to \beta^{4D} + \left( S(\mu) -1 \right) \tilde{\beta} \; , \label{beta4d}
\end{equation}
where $\tilde{\beta}$ is a generic contribution from a single KK level, and where its coefficient is not a constant but instead $S(\mu) = \mu R$, with $\mu^{Max} = \Lambda$, reflecting the power law running behaviour. As a result of faster running, the gauge couplings tend to lower the unification scale down to a relatively low order, which might be accessible to collider experiments.

\par The first version of this model we shall consider, the bulk UED model has the 5-dimensional KK expansions of the weak doublet ($F$) and singlet ($f$) as well as the Higgs and gauge fields ($G$) as shown (the corresponding coupling constants among the KK modes are simply equal to the SM couplings up to normalisation factors, e.g. $\displaystyle Y_U = {Y_U^5}/{\sqrt{\pi R}}$) below:
\begin{eqnarray}
G(x,y) &=& \frac{1}{{\sqrt {\pi R} }}\left\{ G^0(x) + \sqrt 2 \sum\limits_{n = 1}^ \propto  {{G_n}(x)\cos \left(\frac{{ny}}{R}\right)} \right\} \; , \nonumber \\
f(x,y) &=& \frac{1}{{\sqrt {\pi R} }}\left\{ {f_R}(x) + \sqrt 2 \sum\limits_{n = 1}^\infty  \left[f_R^n (x)\cos \left(\frac{{ny}}{R}\right) + f_L^n(x) \sin \left(\frac{{ny}}{R}\right)\right]\right\} \; ,\nonumber \\
F(x,y) &=& \frac{1}{{\sqrt {\pi R} }}\left\{ {F_L}(x) + \sqrt 2 \sum\limits_{n = 1}^\infty  \left[F_L^n (x)\cos \left(\frac{{ny}}{R}\right) + F_R^n(x) \sin \left(\frac{{ny}}{R}\right)\right]\right\} \; \nonumber  . \label{UED7}
\end{eqnarray}
The zero modes in the above equations are identified with the 4-dimensional SM fields, whilst the complex scalar field $H$ and the gauge field $A_\mu$ are $Z_2$ even fields, and there is a left-handed and a right-handed KK mode for each SM chiral fermion. Note that in models with UED momentum conservation in the extra dimensions, we are led to the conservation of KK number at each vertex in the interactions of the 4-dimensional effective theory (or strictly speaking, the KK parity $(-1)^n$ is what remains conserved, where $n$ is the KK number). In the bulk we have the fermion and gauge field interactions as follows:
\begin{eqnarray}
{\cal L}_{Leptons} &=& \int\limits_0^{\pi R} {dy} \{ i\bar L(x,y){\Gamma ^M}{{\cal D}_M}L(x,y) + i\bar e(x,y){\Gamma ^M}{{\cal D}_M}e(x,y)\}
\; ,\nonumber \\
{\cal L}_{Quarks} &=& \int\limits_0^{\pi R} {dy} \{ i\bar Q(x,y){\Gamma ^M}{{\cal D}_M}Q(x,y) + i\bar u(x,y){\Gamma ^M}{{\cal D}_M}u(x,y)\nonumber \\
&+& i\bar d(x,y){\Gamma ^M}{{\cal D}_M}d(x,y)\} \; , \nonumber \\ && \label{UED8}
\end{eqnarray}
where $\Gamma^M=(\gamma^{\mu},i\gamma^5)$, and $M=0,1,2,3,5$, for further details see \cite{Cornell:2010sz}. After integrating out the compactified dimension, the 4-dimensional effective Lagrangian has interactions involving the zero mode and the KK modes. However, these KK modes cannot affect electroweak processes at tree level, and only contribute to higher order electroweak processes. The one-loop Feynman diagram contributions to the Yukawa couplings in the SM and UED model have been explicitly illustrated in \cite{Bhattacharyya:2006ym, Cornell:2010sz, Liu:2011gr, Cheng:1973nv}. In the UED model, where for each energy level $n_i$, we effectively have a heavier duplicate copy of the entire SM particle content. However, new contributions from the $A_5$,
\begin{equation}
A_5(x,y) = \sqrt{\frac{2}{\pi R}} \sum\limits_{n = 1}^\infty A_5^n(x)\sin\left(\frac{ny}{R}\right)  \; , \label{eqn:UED10}
\end{equation}
interactions (that of the fifth component of the vector fields) also contribute. In contrast, the fifth component of the gauge bosons $A_5 (x,y)$ is a real scalar and does not have any zero mode, transforming in the adjoint representation of the gauge group.

\par A simple alternative to this model is that of the brane localised UED model, where we have the same fields but where the fermion matter fields cannot propagate in the bulk and they are restricted to the brane. For the case of brane localised matter fields, only the boson fields (the gauge fields and the scalar fields) can propagate in the bulk space. However, if the compactification radius $R$ is sufficiently large, due to the power law running of the gauge couplings, it will enable us to bring the unification scale down
to an exportable range at the LHC scale.

%
%

\section{The 5D MSSM}\label{sec:3}

\par Another useful model we shall consider is the 5D MSSM defined in \cite{Deandrea:2006mh, Bouchart:2011va, Flacke:2003ac, Hebecker:2001ke, Arkani-Hamed:2001tb, Marcus:1983wb, Mirabelli:1997aj, Buchbinder:2003qu, Cornell:2011fw, ahmad, DDG, DDG1, Antoniadis:1990ew, Csaki:2004ay, Kubyshin:2001mc, Rubakov:2001kp, Perez-Lorenzana:2005iv, Quiros:2003gg, ArkaniHamed:2001tb}. The 5D MSSM is a five dimensional $\mathcal{N}=1$ supersymmetric model compactified on the $S_1/{Z_2}$ orbifold which breaks the 5D Lorentz invariance to the usual 4D one. This breaking gives a momentum conservation along the fifth dimension which conserves the KK number at tree level and KK parity at loop level \cite{Delgado:1999sv, Antoniadis:1992fh, Delgado:1998qr}. One of the main implications of KK-parity invariance is that the lightest KK mode is stable and can be a cold dark matter candidate. In this compactification we can recover the MSSM as zero modes since we obtain chiral fermions. 

\par The gauge sector is then described by a 5D $\mathcal{N}=1$ vector supermultiplet which consists (on-shell) of a 5D vector field $A^M$, a real scalar $S$ and two gauginos, $\lambda$ and $\lambda'$. The action for which can be given by:
\begin{eqnarray}
S_g &=& \int \mathrm{d}^5x\frac{1}{2kg^2}\mathrm{Tr}\left[-\frac{1}{2}F^{MN}F_{MN}-D^MSD_MS-i\overline{\lambda}\Gamma^MD_M
\lambda \right.\nonumber \\
&&\hspace{1cm} -\left. i\overline{\lambda}'\Gamma^MD_M\lambda'+(\overline{\lambda}+\overline{\lambda}')[S,\lambda+\lambda']\right]
\; ,
\label{eqn:0}
\end{eqnarray}
with $D_M=\partial_M+iA_M$ and $\Gamma^M=(\gamma^\mu,i\gamma^5)$. $F^{MN}=-\frac{i}{g}[D^M,D^N]$ and $k$ normalises the trace over the generators of the gauge groups.

\par From the decomposition of the 5D supercharge (which is a Dirac spinor) into two Majorana-type supercharges, which constitute a $\mathcal{N}=2$ superalgebra in 4D, one can rearrange these fields in terms of a $\mathcal{N}=2$, 4D vector supermultiplet, $\Omega=(V\,,\chi)$:
\begin{itemlist}
\item $V$ : $\mathcal{N}=1$ vector supermultiplet containing $A^{\mu}$ and $\lambda$,
\item $\chi$ : $\mathcal{N}=1$ chiral supermultiplet containing $\lambda'$ and $S'=S+iA^5$.
\end{itemlist}
\noindent Both $V$ and $\chi$ (and their component fields) are in the adjoint representation of the gauge group $\mathcal{G}$. Using the supermultiplets one can write the original 5D $\mathcal{N}=1$ supersymmetric action Eq.(\ref{eqn:0}) in terms of $\mathcal{N}=1$ 4D superfields and the covariant derivative in the $y$ direction\cite{Hebecker:2001ke}:
\begin{equation}
S_g = \int\mathrm{d}^5x\mathrm{d}^2\theta\mathrm{d}^2\overline{\theta}\frac{1}{4kg^2}\mathrm{Tr} \left[\frac{1}{4}(W^{\alpha}W_\alpha
\delta(\overline{\theta}^2)+\,h.c)+(e^{-2gV}\nabla_ye^{2gV})^2\right] \; , \label{eqn:0a}
\end{equation}
with $W^{\alpha}=-\frac{1}{4}\overline{D}^2e^{-2gV}D_{\alpha} e^{2gV}$. $D_{\alpha}$ is the covariant derivative in the 4D $\mathcal{N}=1$ superspace (see Refs\cite{Westbook, WessBagger}.) and $\nabla_y=\partial_y+\chi$. To find the Feynman rules to a given order in the gauge coupling $g$, one can expand and quantise the action\cite{Deandrea:2006mh}. The beta functions for the couplings of the operators in the superpotential are governed by the wave function renormalisation constants $Z_{ij} = 1 + \delta Z_{ij}$ due to the non-renormalisation theorem \cite{Iliopoulos:1974zv, Wess:1973kz}. 

\par The Higgs superfields and gauge superfields will always propagate into the fifth dimension. Different possibilities for the matter superfields will be discussed, where superfields containing SM fermions can propagate in the bulk or are restricted to the brane. For the case where all fields can propagate in the bulk, the action for the matter fields would be\cite{Deandrea:2006mh}:
\begin{eqnarray}
S_{matter} &=& \int\mbox{d}^8z\mbox{d}y\left\{ \bar{\Phi}_i\Phi_i + \Phi^c_i\bar{\Phi}^c_i + \Phi^c_i\partial_5\Phi_i \delta(\bar{\theta}) -
\bar{\Phi}_i\partial_5\overline{\Phi}_i^c\delta(\theta) \right. \nonumber \\
& & \left. \hspace{1.5cm} + \tilde{g}(2\bar{\Phi}_iV\Phi_i - 2\Phi_i^cV\bar{\Phi}^c_i + \Phi^c_i\chi \Phi_i\delta(\bar{\theta}) + \bar{\Phi}_i
\bar{\chi}\bar{\Phi}^c_i\delta(\theta)) \right\} \; .
\end{eqnarray}
Again, this action can be expanded and quantised. The $\chi$-field should be odd under $Z_2$ symmetry because it appears together with a derivative $\partial_y$, whereas V is even. For the two matter superfields, we choose $\Phi$ to be even and the conjugate $\Phi^c$ to be odd such that $\Phi^c$ vanishes on the brane. Only the even fields have zero modes.

\par We can write the action for the second case where all superfields containing SM fermions are restricted to the brane. In which case the part of the action involving only gauge and Higgs fields is not modified, whereas the action for the superfields containing the SM fermions becomes:
\begin{equation}
S_{matter} = \int\mbox{d}^8z\mbox{d}y \delta(y) \left\{ \bar{\Phi}_i\Phi_i + 2\tilde{g}\bar{\Phi}_iV\Phi_i \right\} \; .\label{matterbrane}
\end{equation}
Due to the 5D $\mathcal{N}=1$ SUSY, Yukawa couplings are forbidden in the bulk. However, they can be introduced on the branes, which are 4D subspaces with reduced SUSY. One can also add the effective neutrino mass operator (also called lepton number violating Weinberg operator), with dimensional coupling $\tilde{k}_{ij}$ in which we are interested to show its evolution and therefore the Majorana mass term for neutrinos. We will write the following interaction terms, called brane interactions, containing Yukawa-type couplings:
\begin{equation}
S_{brane}=\int\mbox{d}^8z\mbox{d}y\delta(y)\left\{ \left(\frac{1}{6} \tilde{\lambda}_{ijk}\Phi_i\Phi_j\Phi_k - \frac{\tilde{k}_{ij}}{4 M}
L_iH_uL_jH_u\right)\delta(\bar{\theta}) + \mbox{h.c.} \right\} \; , \label{Sbrane}
\end{equation}
where $L$ and $H^{u}$ are the lepton and up-type Higgs doublet chiral superfields respectively. This operator is used to study neutrino masses and mixings, where RGEs for this effective operator have been derived in the context of the four-dimensional SM \cite{Chankowski:1993tx} and MSSM\cite{Antusch:2001ck} and shall be discussed further in section \ref{sec:9}. 

%
%

\section{Gauge couplings}\label{sec:4}

\par The evolution of the gauge couplings in four dimension at one loop are given by:
\begin{eqnarray}
16{\pi ^2}\frac{{d{g_i}}}{{dt}} &=& {b_i}{g_i}^3 \; , \label{gaugecoupling1}
\end{eqnarray}
where $b_i^{SM}=(\frac{{41}}{10},-\frac{{19}}{6},-7)$ and  $b_i^{MSSM}=(\frac{{33}}{5},1,-3)$\cite{Babu:1987im}, using a $SU(5)$ normalisation. If we consider our 5D theory Eq.(\ref{beta4d})can be written in terms of the scale parameter $t$:
\begin{eqnarray}
16{\pi ^2}\frac{{d{g_i}}}{{dt}} &=& [{b_i} + (S(t) - 1){{\tilde b}_i}]{g_i}^3 \; , \label{gaugecoupling3}
\end{eqnarray}
where ${{\tilde b}_i}$ take the following form in the case of the model UED SM\cite{Cornell:2010sz,Liu:2012me}:
\begin{eqnarray}
({\tilde b}_1,{\tilde b}_2, {\tilde b}_3) &=& \left( \frac{1}{10}, - \frac{41}{6}, - \frac{21}{2} \right) + \frac{8}{3} \eta \; , \label{btilde1}
\end{eqnarray}
with $\eta$ being the number of generations of matter fields in the bulk.

\par Next we consider the beta functions of the gauge couplings in the 5D MSSM. In fact, after compactification of the 5D MSSM, where the master beta functions of the gauge couplings in the 5D MSSM as follows \cite{Cornell:2011fw}:
\begin{eqnarray}
({\tilde b}_1,{\tilde b}_2, {\tilde b}_3) &=& \left( \frac{6}{5}, - 2, - 6 \right) + 4\eta \; , \label{btilde4}
\end{eqnarray}
where $\eta$ again represents the number of generations of fermions which propagate in the bulk. 

\par
\begin{figure}[pb]
\begin{center}
\includegraphics[width=0.45\textwidth]{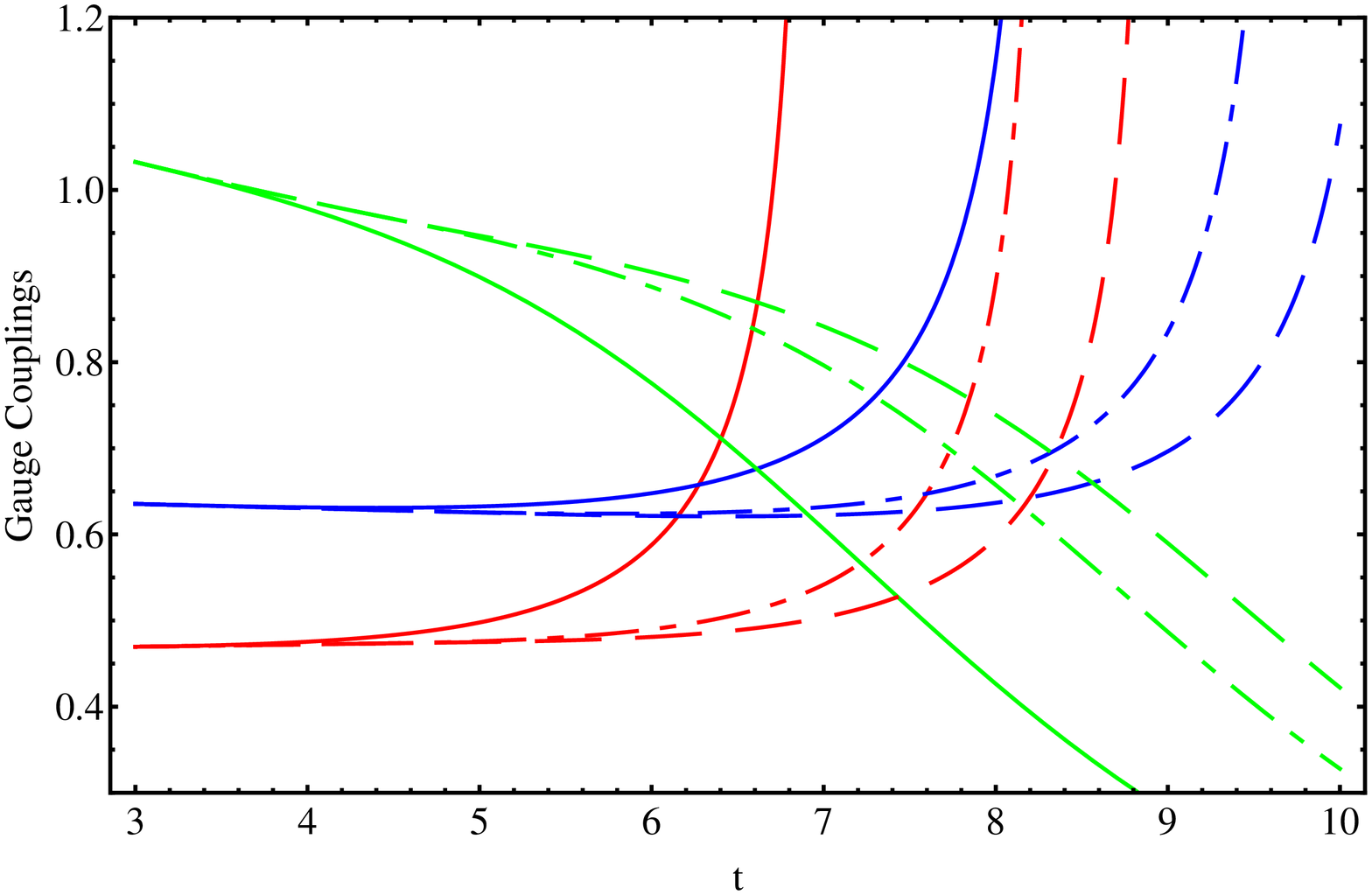}\hspace{0.05\textwidth}
\includegraphics[width=0.45\textwidth]{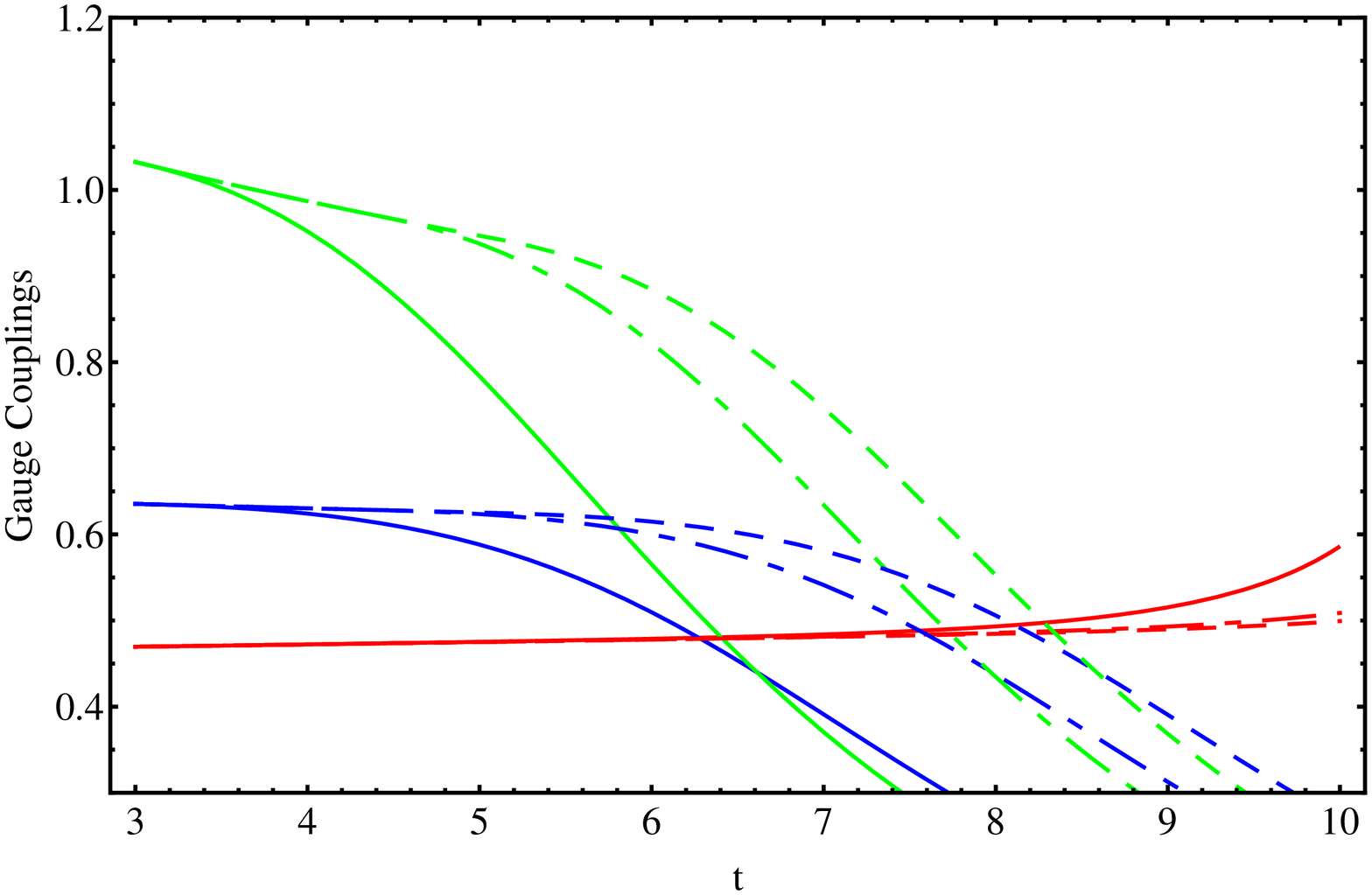}
\vspace*{8pt}
\caption{Gauge couplings ($g_1$ (red), $g_2$ (blue), $g_3$ (green) with: in the left panel, all matter fields in the bulk (UED bulk); and the right panel for all matter fields on the brane (UED brane); for three different values of the compactification scales (2 TeV (solid line), 8 TeV (dot-dashed line), 15 TeV (dashed line)) as a function of the scale parameter $t$ in the UED SM.}
\label{gauge_uedsm}
\end{center}
\end{figure}

\begin{figure}[pb]
\begin{center}
\includegraphics[width=0.45\textwidth]{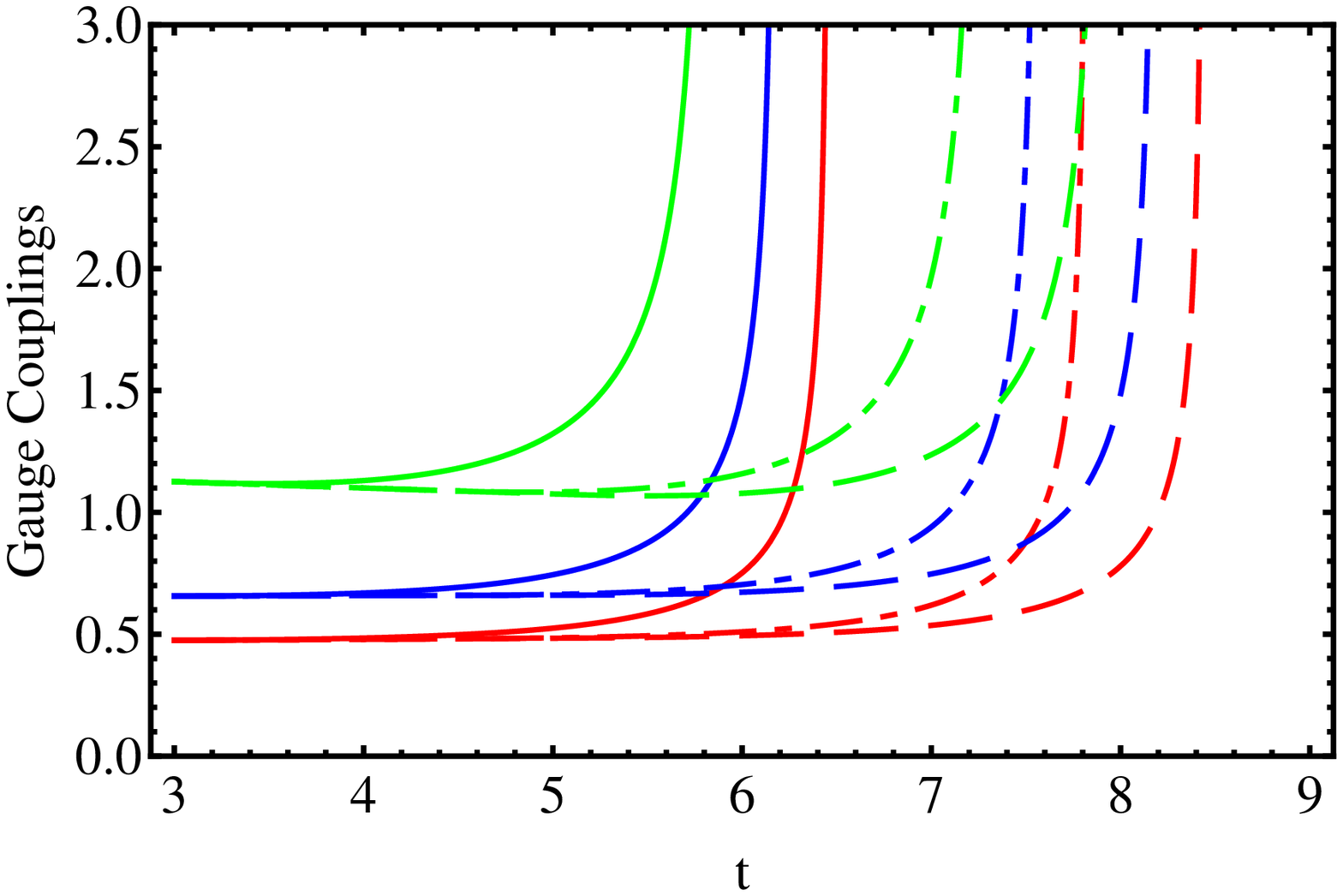}\hspace{0.05\textwidth}
\includegraphics[width=0.45\textwidth]{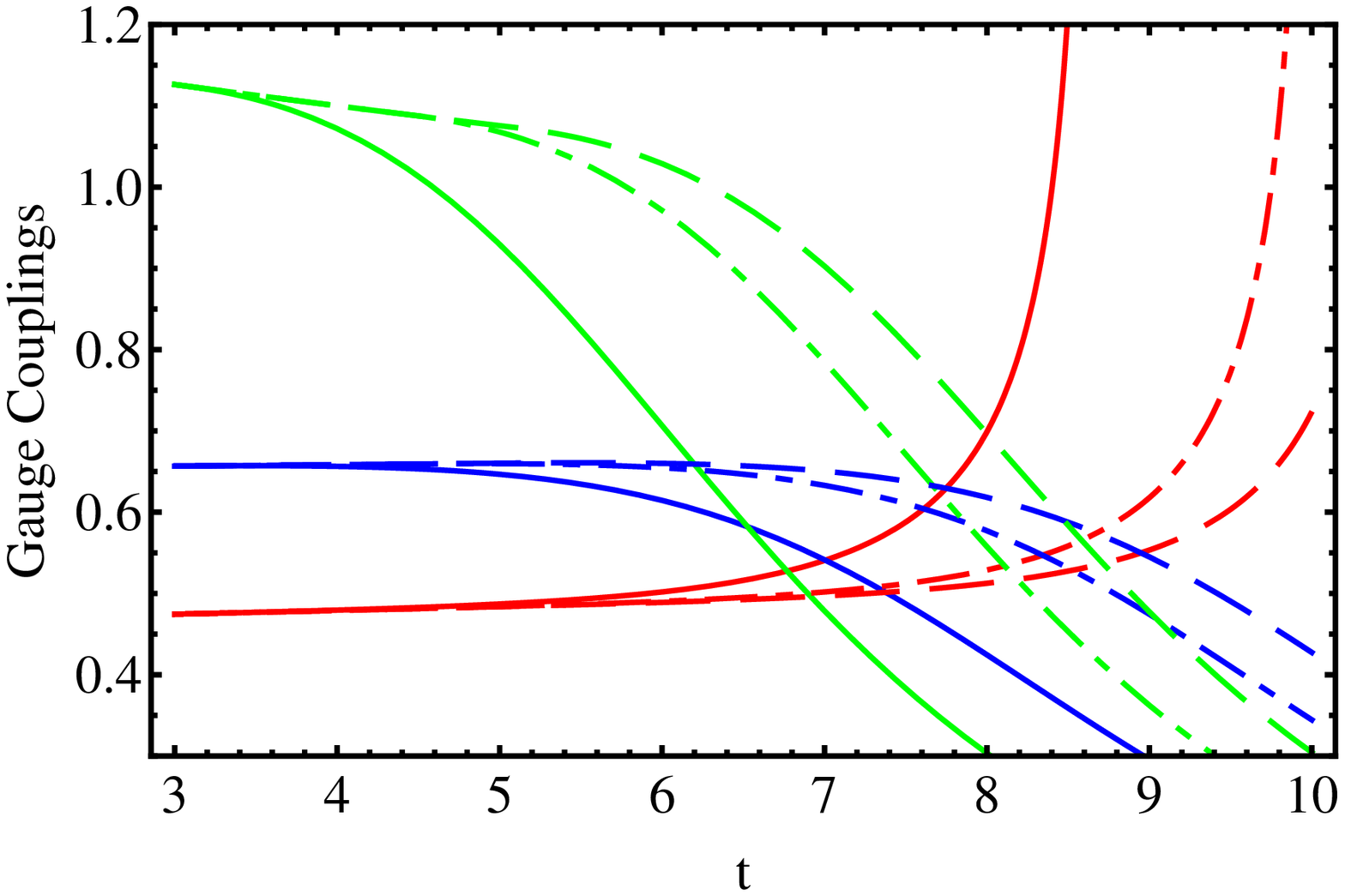}
\vspace*{8pt}
\caption{Gauge couplings ($g_1$ (red), $g_2$ (blue), $g_3$ (green)) with: in the left panel, all matter fields in the bulk; and the right panel for all matter fields on the brane; for three different values of the compactification scales (2 TeV (solid line), 8 TeV (dot-dashed line), 15 TeV (dashed line)) as a function of the scale parameter $t$ in the 5D MSSM.}
\label{gauge_5D}
\end{center}
\end{figure}

 In Figs.\ref{gauge_uedsm} and \ref{gauge_5D} we have plotted the running of the gauge couplings for the UED SM case and the 5D MSSM respectively for the brane localised and bulk field cases, and for several choices of compactification scales for the extra-dimension ($R$). From these plots, and the discussion given in refs. \cite{Bando:2000jd, Bando:2000it}, we find that for the three gauge coupling constants to approach a small region at some value of $t$ requires an extremely large value of $1/R$, which is of no phenomenological interest at present. For the case of our fields being brane localised in the UED model, we see a similar behaviour: the extra-dimensions naturally lead to gauge coupling unification at an intermediate mass scale for the compactification radii considered here. 


%
%

\section{Yukawa evolutions}\label{sec:5}

\par In the quark sector of the SM, we have ten experimentally measurable parameters, i.e. six quark masses, three mixing angles, and one phase (these angles and phase being encoded in the CKM matrix which we shall discuss in section \ref{sec:7}). At present there has been considerable effort to understand the hierarchies of these mixing angles and fermion masses in terms of the RGEs\cite{Cheng:1973nv, Babu:1987im, Sasaki:1986jv, Machacek:1983fi, Liu:2009vh, Balzereit:1998id, Kuo:2005jt}. Note though that when using the RGEs as a probe, the initial values we shall adopt are very important, where we shall scale for the gauge couplings and the fermion masses at the $M_Z$ scale are shown in Table~\ref{ta1}.

\begin{table}[ph]
\tbl{Initial values for the gauge couplings, fermion masses and CKM parameters at $M_Z$ scale. Data is taken from Ref. \protect{\cite{Xing:2007fb, Cornell:2011fw, Nakamura:2010zzi}}.}
{\begin{tabular}{cc|cc} \toprule
Parameter & Value & Parameter & Value \\ \hline
$\alpha_1$ & $0.01696$ &$m_e$ & $0.48657$ MeV \\
$\alpha_2$ & $0.03377$ &$m_\mu$ & $102.718$ MeV \\
$\alpha_3$ & $0.1184$ &$m_\tau$ & $1746.24$ MeV \\
$m_u$ & $1.27$ MeV &$|V_{ub}|$ & $0.00347$ \\
$m_c$ & $0.619$ GeV &$|V_{cb}|$ & $0.0410$ \\
$m_t$ & $171.7$ GeV &$|V_{us}|$ & $0.2253$ \\
$m_d$ & $2.90$ MeV &$J$ & $2.91 \times 10^{-5}$ \\
$m_s$ & $55$ MeV &&\\
$m_b$ & $2.89$ GeV &&
\end{tabular}\label{ta1}}
\end{table}

\par Furthermore, we shall also attempt, in section \ref{sec:9}, to develop the RGEs of the lepton sector (including possible mixing angles and phases), which will require knowledge of the evolution of a parameter $k$, where the lowest order operator which generates Majorana neutrino masses after electroweak symmetry breaking (EWSB), is the lepton-number violating Weinberg operator \cite{Weinberg:1979sa}. This lowest order operator (appearing with dimension $d=5$ in four space-time dimensions) can be written as:
\begin{equation}
-\frac{\tilde{k}_{ij}}{4M}(\bar{L}^{ci}_{\alpha} \epsilon^{\alpha \beta} \phi_{\beta})(L^{j}_{\delta} \epsilon^{\delta \gamma} \phi_{\gamma})
+h.c. \; ,\label{eq:5dop}
\end{equation}
where $L$ and $\phi$ are the lepton and the Higgs doublet fields. $M$ is the typical heavy energy scale for the range of validity of the low-energy effective theory, where renormalisation group equations for this effective operator have been derived in the context of the four-dimensional SM and MSSM\cite{Chankowski:1993tx,Antusch:2001ck}.

\par In the present case we consider the effective neutrino mass operator with dimensional coupling $\tilde{k}_{ij}$; after spontaneous symmetry breaking, the Majorana neutrino masses can be written as $m_\nu \equiv k v^2 sin^2 \beta$ ($v$ being the vev of the Higgs field and $\tan \beta$, the ratio of the vevs of our two Higgs doublets) and $k=\tilde{k}_{ij}/(2 M \pi R)$ for bulk propagating, and $k=\tilde{k}/(2M)$ for brane localised matter superfield scenarios respectively.

\par As such, we have set ${M_Z}$ as the renormalisation point, and use $t=\ln (\frac{\mu}{M_Z})$ and $S(t) = {e^t}{M_Z}R$. The general form of evolution equations for Yukawa couplings and neutrino $k$ coupling at the one loop can be written in the following form Refs\cite{Blennow:2011mp,yukawaSM, yukawaSM1, Antusch:2003kp}:
\begin{eqnarray}
16{\pi ^2}\frac{{d{Y_d}}}{{dt}} &=& {Y_d}\left\{ {{T_t}C_1 - {G_d} + \frac{3}{2}(Y_d^\dag {Y_d} - Y_u^\dag {Y_u})C_2} \right\} \; , \nonumber
\\
16{\pi ^2}\frac{{d{Y_u}}}{{dt}} &=& {Y_u}\left\{ {{T_t}C_1 - {G_u} + \frac{3}{2}(Y_u^\dag {Y_u} - Y_d^\dag {Y_d})C_2} \right\} \; ,
\label{eqn:Yukawa} \\
16{\pi ^2}\frac{{d{Y_e}}}{{dt}} &=& {Y_e}\left\{ {{T_t}C_1 - {G_e} + \frac{3}{2}(Y_e^\dag {Y_e})C_2} \right\} \; , \nonumber \\
16{\pi ^2}\frac{{d{k}}}{{dt}} &=& \alpha k + \left( [{Y_e}^T {Y_e}^*]k+k[{Y_e}^\dag Y_e] \right)C_3 \; . \nonumber
\end{eqnarray}
where ${T_t}= Tr \Big[ 3Y_d^\dag {Y_d} + 3Y_u^\dag {Y_u} + Y_e^\dag {Y_e} \Big]$.

%

\subsection{Standard Model}\label{SMcoef}

\par The SM is a limiting case for the UED, where the KK states decouple. The coefficients in the evolution equation are defined by: ${G_d}_{SM}=( {\frac{1}{{4}}g_1^2 + \frac{9}{4}g_2^2 + 8g_3^2} )$, ${G_u}_{SM}=( {\frac{{17}}{{20}}g_1^2 + \frac{9}{4}g_2^2 + 8g_3^2} )$, ${G_e}_{SM}=( {\frac{9}{4}g_1^2 + \frac{9}{4}g_2^2} )$, ${\alpha}_{SM}=2\; T_t - 3g_2^2 +\lambda$, ${C_1}_{SM}=1$, ${C_2}_{SM}=1$, ${C_3}_{SM}=-\frac{3}{2}$. 

%

\subsection{UED SM Bulk}\label{UEDBulkcoef}

\par The UED contribution is obtained when KK states enter, where due to the orbifolding the zero mode for fermions are chiral, which are replaced by Dirac fermions at each KK level. This lead to the factor 2 appearing in $C_1$ and $C_2$ since the KK left and right-handed chiral states contribute to the closed fermion one loop diagrams. That is, ${G_d}_{UEDBulk}=( {\frac{17}{{120}}g_1^2 + \frac{15}{8}g_2^2 + \frac{28}{3}g_3^2} )(S(t)-1)$, ${G_u}_{UEDBulk}=( {\frac{{101}}{{120}}g_1^2 + \frac{15}{8}g_2^2 + \frac{28}{3}g_3^2} )(S(t)-1)$, ${G_e}_{UEDBulk}=( {\frac{99}{40}g_1^2 + \frac{15}{8}g_2^2} )(S(t)-1)$, ${\alpha}_{UEDBulk}=(S(t)-1)\Big( 4T_t -\frac{3}{20} g_1^2-\frac{11}{4}g_2^2 +\lambda \Big)$, ${C_1}_{UEDBulk}=2(S(t)-1)$, ${C_2}_{UEDBulk}=(S(t)-1)$, ${C_3}_{UEDBulk}=(S(t)-1)$. 

%

\subsection{UED SM Brane}\label{UEDBranecoef}

\par For the case where the fermions are restricted to the brane, we obtain the coefficients from Ref\cite{Liu:2012me}. and Ref \cite{Blennow:2011mp}. ${G_d}_{UEDBrane}=( {\frac{1}{{4}}g_1^2 + \frac{9}{4}g_2^2 + 8g_3^2} )2(S(t)-1)$, ${G_u}_{UEDBrane}=( {\frac{{17}}{{20}}g_1^2 + \frac{9}{4}g_2^2 +8g_3^2} )2(S(t)-1)$, ${G_e}_{UEDBrane}=( {\frac{9}{4}g_1^2 + \frac{9}{4}g_2^2} )2(S(t)-1)$, ${\alpha}_{UEDBrane}= 2(S(t)-1)\Big(-3g_2^2 +\lambda \Big)$, ${C_1}_{UEDBrane}=0$, ${C_2}_{UEDBrane}=2(S(t)-1)$, ${C_3}_{UEDBrane}=2(S(t)-1)$. Note that the coefficient $C_1 = 0$ since we do not have a trace of fermionic loops as the fermions are restricted to the brane.

%
%

\subsection{Yukawa evolutions in the MSSM and 5D MSSM}\label{MSSMcoef}

\par Similarly, the general form of the evolution equations for the various MSSMs, where we shall use a notation similar to the ones of Refs\cite{Antusch:2003kp,Deandrea:2006mh}.
\begin{eqnarray}
16{\pi ^2}\frac{{d{Y_d}}}{{dt}} &=& {Y_d}\left\{ {{T_d}\tilde{C} - {G_d} + (3Y_d^\dag {Y_d} + Y_u^\dag {Y_u})C} \right\} \; , \nonumber\\
16{\pi ^2}\frac{{d{Y_u}}}{{dt}} &=& {Y_u}\left\{ {{T_u}\tilde{C} - {G_u} + (3Y_u^\dag {Y_u} + Y_d^\dag {Y_d})C} \right\} \; , \\
16{\pi ^2}\frac{{d{Y_e}}}{{dt}} &=& {Y_e}\left\{ {{T_e}\tilde{C} - {G_e} + (3Y_e^\dag {Y_e})C} \right\} \; , \nonumber \\
16{\pi ^2}\frac{{d{k}}}{{dt}} &=& \alpha k + \left( [{Y_e}^T {Y_e}^*]k+k[{Y_e}^\dag Y_e] \right) C \; . \nonumber
\end{eqnarray}
where ${T_d}=3\; Tr(Y_d^\dag {Y_d}) + Tr(Y_e^\dag {Y_e})$, ${T_u}=3\; Tr(Y_u^\dag {Y_u})$, ${T_e}=3\; Tr(Y_d^\dag {Y_d}) + Tr(Y_e^\dag {Y_e})$. Where for the MSSM, as a limiting case of the 5D models we shall consider in the following, and also when $0 < t < \ln (\frac{1}{{{M_Z}R}})$ the coefficients in the evolution equations are: ${G_d}_{MSSM}=( {\frac{7}{{15}}g_1^2 + 3g_2^2 + \frac{{16}}{3}g_3^2} )$, ${G_u}_{MSSM}=( {\frac{{13}}{{15}}g_1^2 + 3g_2^2 + \frac{{16}}{3}g_3^2})$, ${G_e}_{MSSM}( {\frac{9}{5}g_1^2 + 3g_2^2})$, ${\alpha}_{MSSM}=2\; {T_u} - \frac{{6}}{5}g_1^2 - 6g_2^2$, ${C}_{MSSM}=1$, ${\tilde{C}}_{MSSM}= 1$. 

%

\subsection{Bulk MSSM}\label{5Dbulkcoef}

\par The coefficients in the 5D MSSM, for all three generations propagating in the bulk, can be expressed as: ${G_d}_{5Dbulk}=( {\frac{7}{{15}}g_1^2 + 3g_2^2 + \frac{{16}}{3}g_3^2} )S(t)$, ${G_u}_{5Dbulk}( {\frac{{13}}{{15}}g_1^2 + 3g_2^2 + \frac{{16}}{3}g_3^2} )S(t)$, ${G_e}_{5Dbulk}=( {\frac{9}{5}g_1^2 + 3g_2^2} )S(t)$, ${\alpha}_{5Dbulk}=2 \tilde{C}_{5Dbulk}\; {T_u} - (\frac{6}{5} g_1^2 + 6 g_2^2) S(t)$, ${C}_{5Dbulk}=\pi S(t)^2$, ${\tilde{C}}_{5Dbulk}=\pi S(t)^2$.

%

\subsection{Brane MSSM}\label{5Dbranecoef}

\par However, when all matter superfields are constrained to live on the 4D brane, the coefficients of the evolution equations are given by: ${G_d}_{5Dbrane}=( {\frac{19}{{30}}g_1^2 + \frac{9}{{2}}g_2^2 + \frac{{32}}{3}g_3^2} )S(t)$, ${G_u}_{5Dbrane}=( {\frac{{43}}{{30}}g_1^2 + \frac{9}{{2}}g_2^2 + \frac{{32}}{3}g_3^2} )S(t)$, ${G_e}_{5Dbrane}=( {\frac{33}{10}g_1^2 +\frac{9}{{2}}g_2^2} )S(t)$, ${\alpha}_{5Dbrane}=2\; {T_u} - (\frac{9}{5} g_1^2 + 9 g_2^2) S(t)$, ${C}_{5Dbrane}=2 S(t)$, ${\tilde{C}}_{5Dbrane}=1$.

%
%

\section{Scaling of the Yukawa couplings and the CKM matrix}\label{sec:7}

\par It is well known that in the SM, the quark sector's flavor mixing is parameterised by the CKM matrix, which makes it possible to explain all flavor changing weak decay processes and CP-violating phenomena to date. In particular, for the standard parameterisation of the CKM matrix, which has the form:
\begin{eqnarray}
V_{CKM} = \left( {\begin{array}{ccc}
{{c_{12}}{c_{13}}}&{{s_{12}}{c_{13}}}&{{s_{13}}{e^{ - i{\delta}}}}\\
{ - {s_{12}}{c_{23}} - {c_{12}}{s_{23}}{s_{13}}{e^{i{\delta}}}}&{{c_{12}}{c_{23}} - {s_{12}}{s_{23}}{s_{13}}{e^{i{\delta}}}}&{{s_{23}}{c_{13}}}\\
{{s_{12}}{s_{23}} - {c_{12}}{c_{23}}{s_{13}}{e^{i{\delta}}}}&{ - {c_{12}}{s_{23}} - {s_{12}}{c_{23}}{s_{13}}{e^{i{\delta}}}}&{{c_{23}}{c_{13}}}
\end{array}} \right) \; , \label{CKM2}
\end{eqnarray}
where $s_{12} = \sin\theta_{12}$, $c_{12} = \cos\theta_{12}$ etc. are the sines and cosines of the three mixing angles $\theta_{12}$, $
\theta_{23}$ and $\theta_{13}$, and $\delta$ is the CP violating phase.

\par The CKM matrix arises from a consideration of the square of the quark Yukawa coupling matrices being diagonalised by using two unitary matrices $U$ and $V$,
\begin{eqnarray}
\mathrm{diag}\left(f_u^2,f_c^2,f_t^2\right)&=& U Y_u^\dag {Y_u}{U^\dag }\; ,  \nonumber\\
\mathrm{diag}\left(h_d^2,h_s^2,h_b^2\right)&=& V Y_d^\dag {Y_d}{V^\dag }\; , \label{eqn:341}
\end{eqnarray}
in which $f_u^2$, $f_c^2$, $f_t^2$ and $h_d^2$, $h_s^2$, $h_b^2$ are the eigenvalues of $Y_u^\dag Y_u$ and $ Y_d^\dag Y_d$ respectively, with ${V_{CKM}} = U{V^\dag }$. From the full set of one-loop coupled RGE for the Yukawa couplings and the CKM matrix, together with those for the gauge coupling equations, one can obtain the renormalisation group flow of all observables related to up- and down-quark masses and the CKM matrix elements.

\par We write down the general form for the evolution of $f_i^2$, $h_j^2$ and the variation of each element of the CKM matrix $V_{ik}$ \cite{Cornell:2010sz, Cornell:2011fw, Liu:2012me} in the SM, the UED SM, the MSSM and the 5D MSSM.

%

\subsection{SM, UED Bulk SM and UED Brane SM}

\begin{eqnarray}
16{\pi ^2}\frac{{df_i^2}}{{dt}} &=& f_i^2[2({T_u}A - {G_u}) + 3{B}f_i^2 - 2{B}\sum\limits_j  {h_j ^2} {\left| {{V_{ij}}} \right|^2}]\; , \nonumber\\
16{\pi ^2}\frac{{dh_j ^2}}{{dt}} &=& h_j ^2[2({T_d}A - {G_d}) + 3{B}h_j ^2 - 2{B}\sum\limits_i {f_i^2} {\left| {{V_{ij}}} \right|^2}] \;,
\label{eqn:343} \\
16{\pi ^2}\frac{{dy_e^2}}{{dt}} &=& y_e^2[2({T_e}A - {G_e}) + 3{B}y_e^2] \; , \nonumber \\
16{\pi ^2}\frac{{d{V_{ik}}}}{{dt}} &=& -\frac{3}{2}{B}\left[\sum\limits_{m,j \ne i} {\frac{{f_i^2 + f_j^2}}{{f_i^2 - f_j^2}}} h_m^2{V_{im}}
V_{jm}^*{V_{jk}} + \sum\limits_{j,m  \ne k} {\frac{{h_k^2 + h_m^2}}{{h_k^2 - h_m^2}}} f_j^2V_{jm}^*{V_{jk}}{V_{im}}\right] \; , \nonumber
\end{eqnarray}
where $A=B=1$ in the SM,  $A=2S(t)-1$, $B=S(t)$ in the UED Bulk SM and $A=0$, $B=2S(t)$ in the UED Brane SM. The gauge couplings $G$ for the SM, the UED Bulk SM and the UED Brane SM are written in sec.(\ref{SMcoef}), sec.(\ref{UEDBulkcoef}) and sec.
(\ref{UEDBranecoef}) respectively.

%

\subsection{MSSM, 5D bulk and 5D brane}

\begin{eqnarray}
16{\pi ^2}\frac{{df_i^2}}{{dt}} &=& f_i^2[2({T_u}\tilde{C} - {G_u}) + 6{C}f_i^2 + 2{C}\sum\limits_j  {h_j ^2} {\left| {{V_{ij}}} \right|^2}]\;
\nonumber\\
16{\pi ^2}\frac{{dh_j ^2}}{{dt}} &=& h_j ^2[2({T_d}\tilde{C} - {G_d}) + 6{C}h_j ^2 + 2{C}\sum\limits_i {f_i^2} {\left| {{V_{ij}}} \right|^2}] \;,
\label{eqn:344} \\
16{\pi ^2}\frac{{dy_e^2}}{{dt}} &=& y_e^2[2({T_e}\tilde{C} - {G_e}) + 6{C}y_e^2] \; , \nonumber \\
16{\pi ^2}\frac{{d{V_{ik}}}}{{dt}} &=& {C}\left[\sum\limits_{m,j \ne i} {\frac{{f_i^2 + f_j^2}}{{f_i^2 - f_j^2}}} h_m^2{V_{im}}V_{jm}^*{V_{jk}} +
\sum\limits_{j,m  \ne k} {\frac{{h_k^2 + h_m^2}}{{h_k^2 - h_m^2}}} f_j^2V_{jm}^*{V_{jk}}{V_{im}}\right] \; , \nonumber
\end{eqnarray}
where we use the same forms as in secs.(\ref{MSSMcoef}, \ref{5Dbulkcoef}, \ref{5Dbranecoef}) to fix the coefficients $C$, $\tilde{C}$ and gauge couplings $G$ to describe each model.

%

\subsection{Top Yukawa coupling}

\subsubsection*{UED SM: Bulk and Brane cases}

\par
\begin{figure}[pb]
\begin{center}
\includegraphics[width=0.45\textwidth]{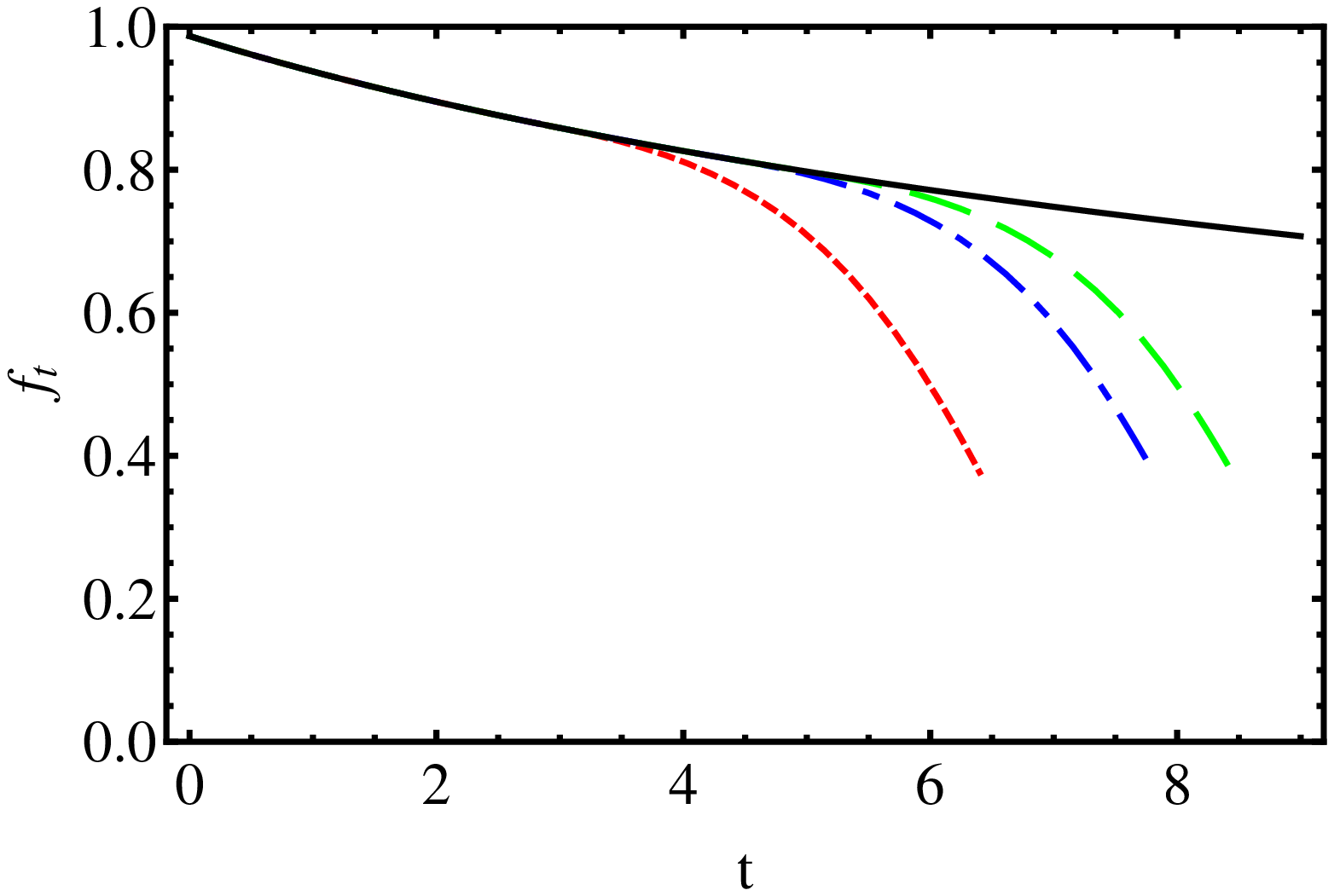}\hspace{0.05\textwidth}
\includegraphics[width=0.45\textwidth]{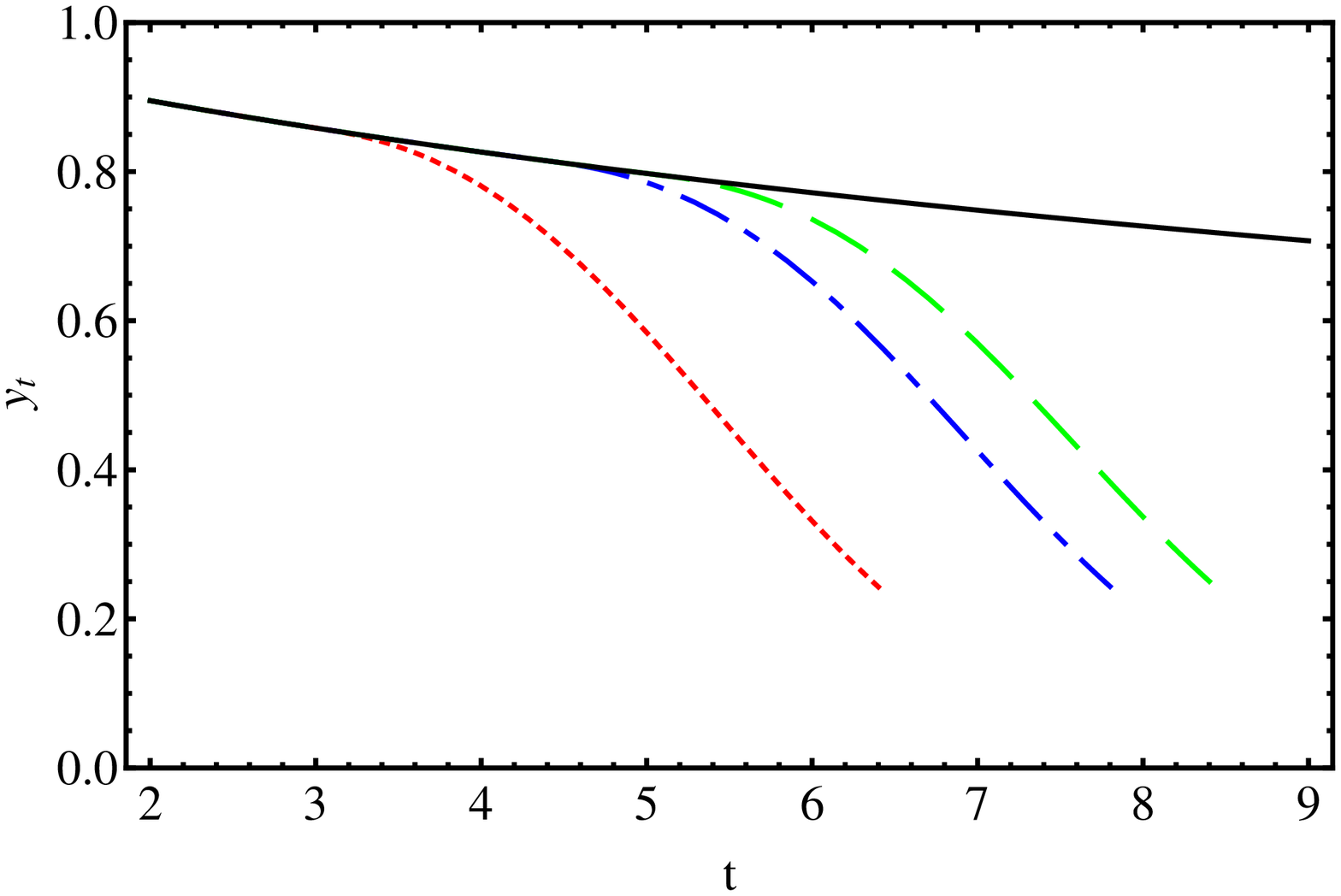}
\vspace*{8pt}
\caption{The Yukawa coupling $f_t$ for the top quark in the UED SM as a function of the scale parameter $t$, for the bulk case (left panel) and the brane case (right panel) where the solid line is the SM evolution and for different compactification scales: $R^{-1}$ = 2 TeV (red, dotted line), 8 TeV (blue,dot-dashed line), and 15 TeV (green,dashed line).}
\label{fig:ft_uedsm}
\end{center}
\end{figure}

 In Fig.\ref{fig:ft_uedsm} the initial Yukawa couplings are given by the ratios of the fermion masses to the Higgs vacuum expectation value. The Yukawa couplings evolve in the usual logarithmic fashion when the energy is below 2 TeV, 8 TeV, and 15 TeV for the three different cases. However, once the first KK threshold is reached, the contributions from the KK states become more and more significant. The evolution of $f_t$ (see Eq.(\ref{eqn:343})) depends explicitly on the cutoff $\Lambda$, which have finite one-loop corrections to the beta functions at each massive KK excitation level. Therefore, the running of the Yukawa couplings, or more precisely, the one-loop KK corrected effective four dimensional Yukawa couplings, begins to deviate from their normal orbits and start to evolve faster and faster. Note also observe that the Yukawa couplings are quickly evolving to zero, however, a satisfactory unification of these seems to still be lacking. As such, we have so far observed the Yukawa couplings all decrease with increasing energy, which agrees with what is observed in the SM, however, the Yukawa couplings are driven dramatically towards extremely weak values at a much faster rate. This is an interesting feature that distinguishes the UED model from that of the SM.

\subsubsection*{5D MSSM Bulk}

\par The 4D MSSM contains the particle spectrum of a two-Higgs doublet model extension of the SM and the corresponding supersymmetric partners. The two Higgs doublets $H_u$ and $H_d$, with opposite hypercharges, are responsible for the generation of the up-type and down-type quarks respectively. The vacuum expectation values of the neutral components of the two Higgs fields satisfy the relation ${v_u}^2 + {v_d}^2 = {\left( {\frac{{246}}{{\sqrt 2 }}} \right)^2} = {\left( {174GeV} \right)^2}$. As a result, the initial Yukawa couplings are given by the ratios of the fermion masses to the appropriate Higgs vacuum expectation values as follows:
\begin{eqnarray}
{f_{u,c,t}} = \frac{{{m_{u,c,t}}}}{{{v_u}}}\;\; , \;\;
{h_{d,s,b}} = \frac{{{m_{d,s,b}}}}{{{v_d}}}\;\; , \;\;
{y_{e,\mu ,\tau }} = \frac{{{m_{e,\mu ,\tau }}}}{{{v_d}}} \; , \label{5DMSSMCKM48}
\end{eqnarray}
where we define $\tan \beta  = v_u/v_d$, which is the ratio of vacuum expectation values of the two Higgs fields $H_u$ and $H_d$.

\par Furthermore, below the supersymmetric breaking scale the Yukawa and gauge couplings run in the usual logarithmic fashion, giving a rather slow change for their values. Therefore, for supersymmetric breaking theories around TeV scales, for simplicity, we take the supersymmetric breaking scale $M_{SUSY}=M_Z$ in the present numerical study, and run the RGEs from $M_Z$ up to the high energy scales for our three different compactification scales.

\par
\begin{figure}[pb]
\begin{center}
\includegraphics[width=0.45\textwidth]{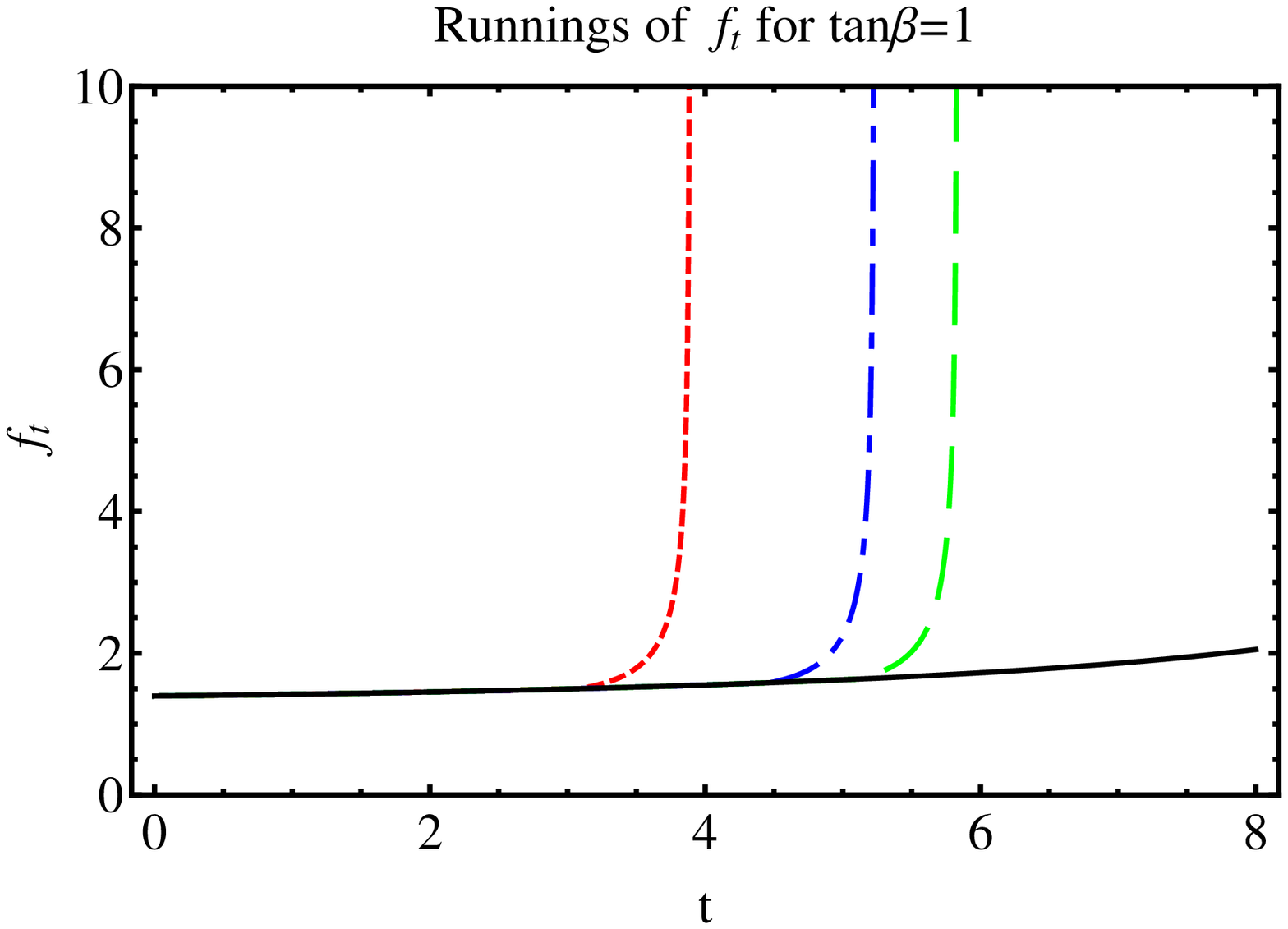}\hspace{0.05\textwidth}
\includegraphics[width=0.45\textwidth]{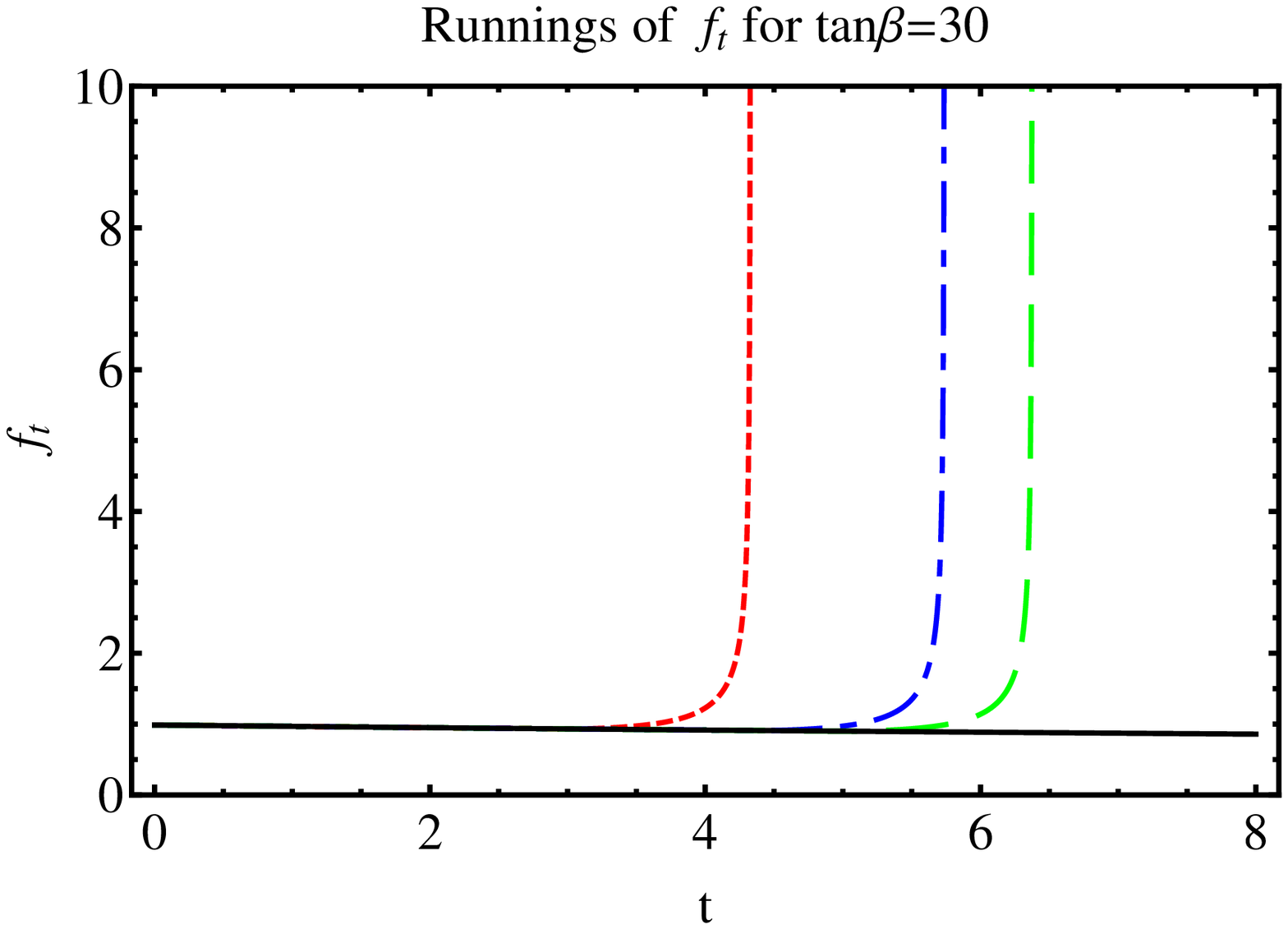}
\vspace*{8pt}
\caption{The Yukawa coupling $f_t$ for the top quark in the bulk case of 5D MSSM as a function of the scale parameter $t$, for (left panel) $\tan\beta=1$ and (right panel) $\tan\beta = 30$ where the solid line is the MSSM evolution and for different compactification scales: $R^{-1}$ = 2 TeV (red, dotted line), 8 TeV (blue, dot-dashed line), and 15 TeV (green, dashed line).}
\label{fig:ft_bulk}
\end{center}
\end{figure}

 Once again, once the first KK threshold is crossed, the power law running of the various beta functions causes the Yukawa coupling to rapidly increase following the rapid increase in the gauge coupling constants. This behaviour can be observed for both small and large $\tan\beta$ cases. However, as illustrated in the first graph of Fig.~\ref{fig:ft_bulk}, for small $\tan\beta$, the Yukawa coupling has a large initial value, therefore it blows up at a relatively low energy as compared with the case for large $\tan\beta$. As a result, as one evolves upward in the scale, the top Yukawa coupling is rising with a fast rate and is pushed up against the Landau pole where it becomes divergent and blows up. In the vicinity of this singular point the perturbative calculation becomes invalid, and the higher order corrections become significant. 

\subsubsection*{5D MSSM Brane}

\par
\begin{figure}[pb]
\begin{center}
\includegraphics[width=0.45\textwidth]{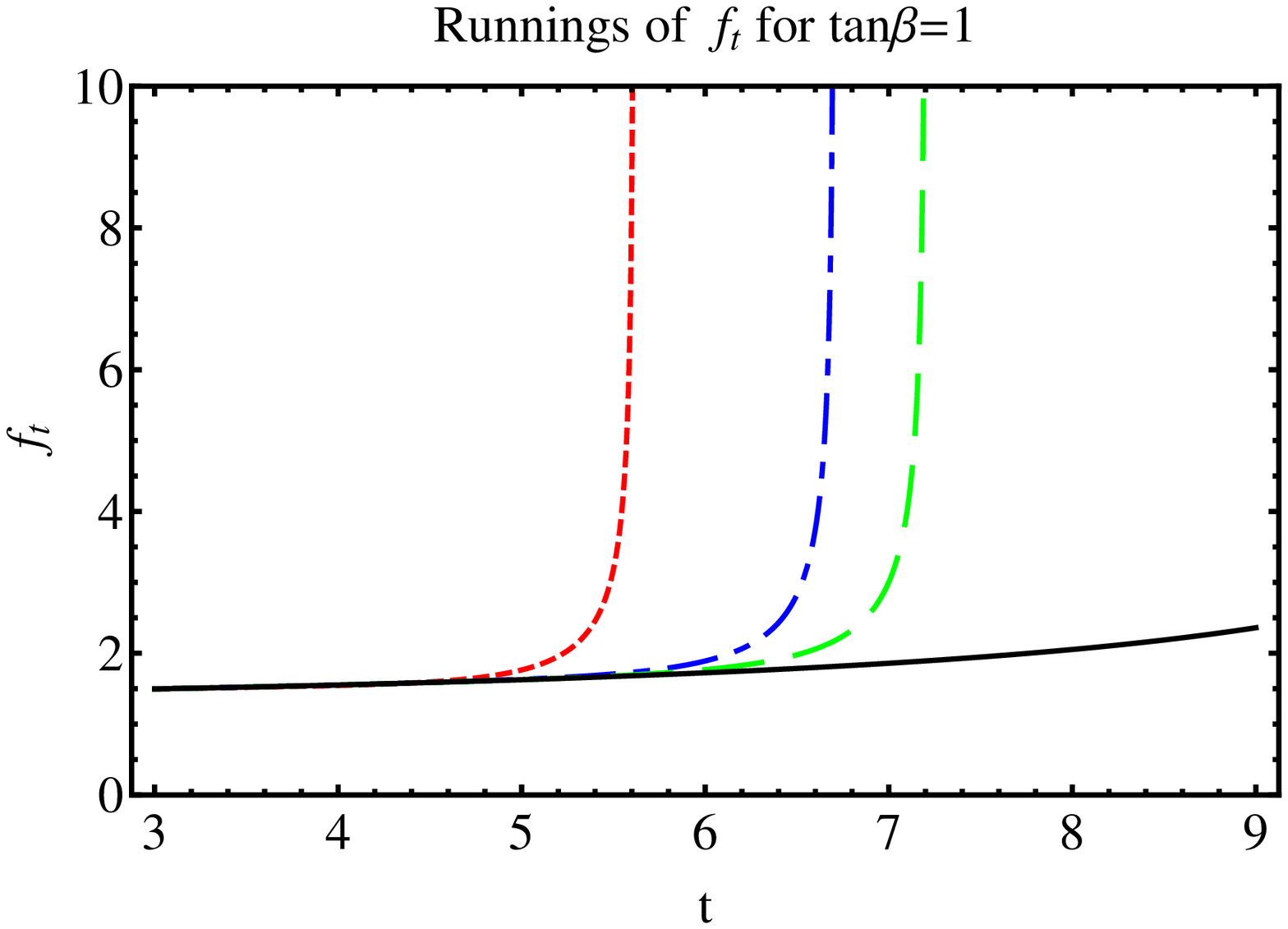}\hspace{0.05\textwidth}
\includegraphics[width= 0.45\textwidth]{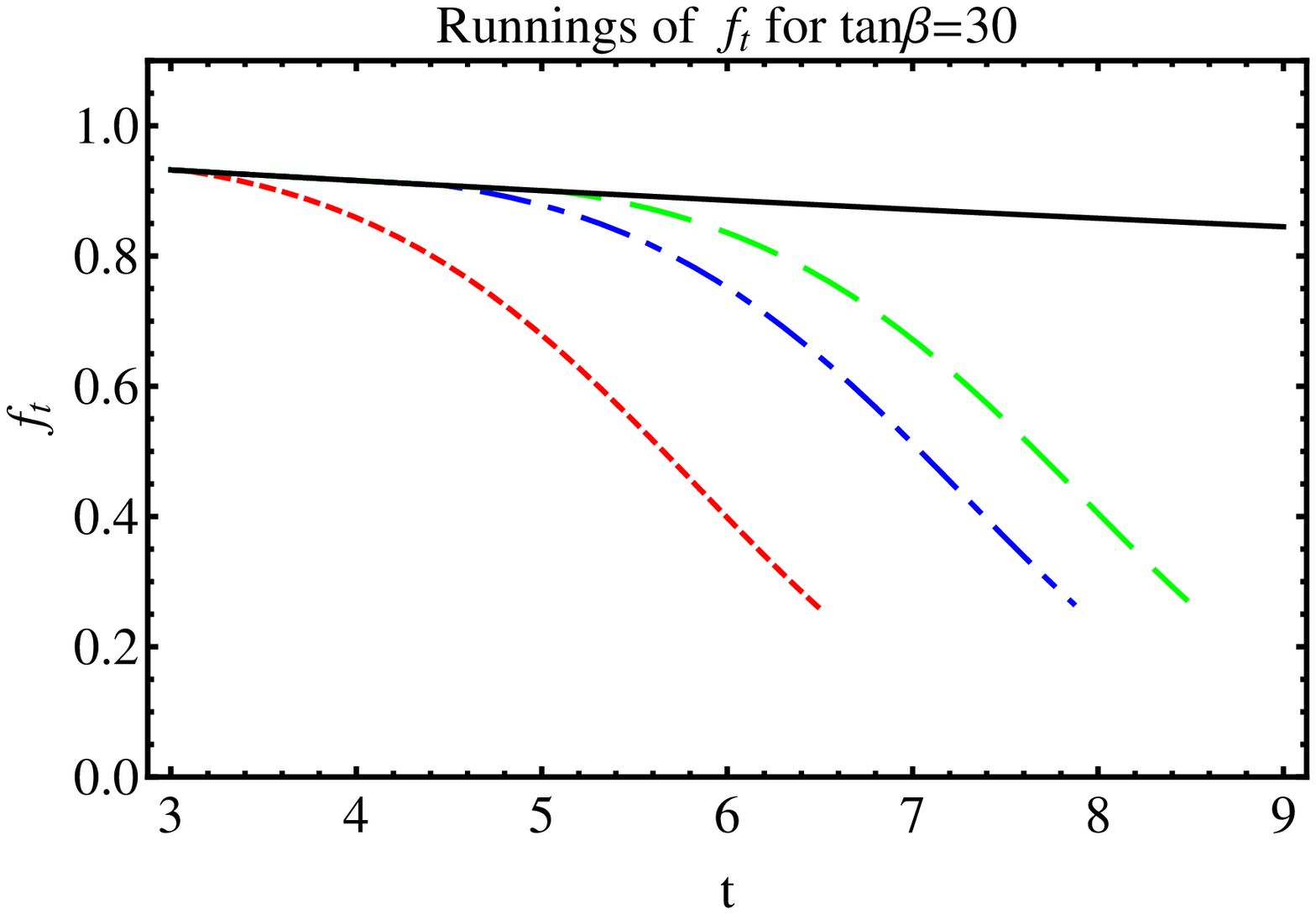}
\vspace*{8pt}
\caption{The Yukawa coupling $f_t$ for the top quark in the brane case of 5D MSSM as a function of the scale parameter $t$, for (left panel) $\tan\beta=1$ and (right panel) $\tan\beta = 30$ where the solid line is the MSSM evolution and for different compactification scales: $R^{-1}$ = 2 TeV (red, dotted line), 8 TeV (blue, dot-dashed line), and 15 TeV (green, dashed line).}
\label{fig:ft_brane}
\end{center}
\end{figure}

 In the brane localised matter field scenario, the beta function has only linear terms in $S(t)$, which is comparable with the $S(t)$ term in the beta function for the gauge couplings. As depicted in Fig.~\ref{fig:ft_brane}, for a small value of $\tan\beta$, we have a large initial value of $f_t$ and the gauge coupling contribution to the Yukawa beta function is sub-dominant only. Therefore  the Yukawa coupling $f_t$ increases rapidly as one crosses the KK threshold, resulting in a rapid approach of the singularity before the unification scale is reached. However, for an intermediate value of $\tan\beta$, we have a relative smaller initial condition for the top Yukawa coupling and the Yukawa terms in the beta function become less important. The contributions from the gauge couplings may then become significant, which leads to a net negative contribution to the beta functions. Therefore, the curvature of the trajectory of the top Yukawa evolution might change direction, becoming more obvious for a large value of $\tan\beta$. This behaviour provides a very clear phenomenological signature, especially for scenarios with a larger $\tan\beta$ and that are valid up to the unification scale.

%

\subsection{CKM Matrix}

\subsubsection*{UED SM}

\par
\begin{figure}[pb]
\begin{center}
\includegraphics[width=0.45\textwidth]{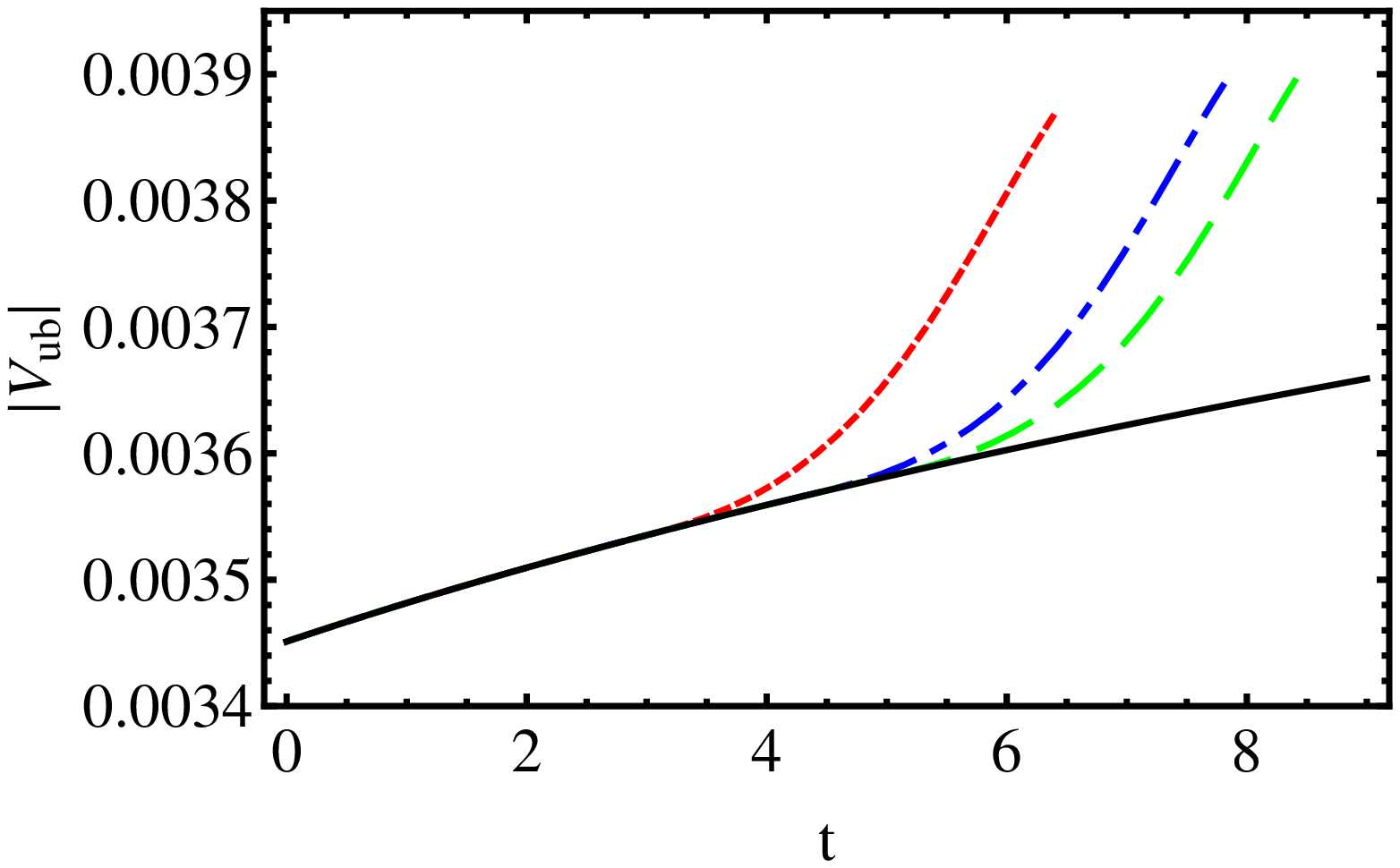}\hspace{0.05\textwidth}
\includegraphics[width=0.45\textwidth]{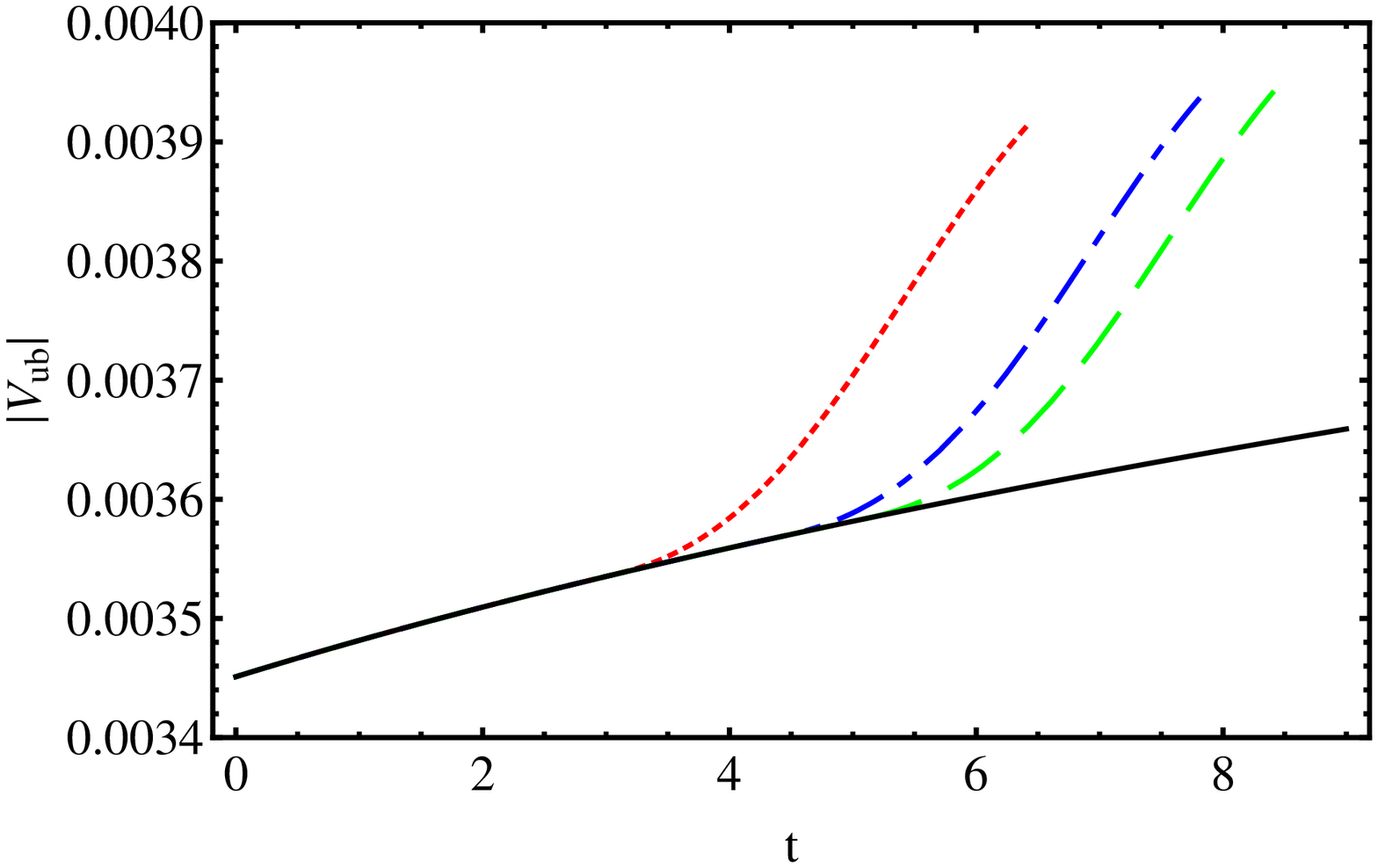}
\vspace*{8pt}
\caption{The CMK matrix elements $|V_{ub}|$ in the UED SM as a function of the scale parameter $t$, for the bulk case (left panel) and the brane case (right panel) where the solid line is the SM, for different compactification scales: $R^{-1}$ = 2 TeV (red, dotted line), 8 TeV (blue,dot-dashed line), and 15 TeV (green, dashed line).}
\label{fig:Vub_uedsm}
\end{center}
\end{figure}

 In Fig.\ref{fig:Vub_uedsm} we plot the evolution of $|V_{ub}|$ for the UED bulk and brane cases. For the evolution of $|V_{cb}|$ and $|V_{us}|$ we can observe similar behaviours, i.e., they all increase with the energy scale; as can be seen from Eq.(\ref{eqn:343}), the evolution of the CKM matrix is governed by the Yukawa couplings and the factor $S(t)$. They evolve faster in the region where the power law scaling of the Yukawa couplings becomes substantial. Therefore, the renormalisation effect is explicit for mixings involving the third family.

\subsubsection*{5D MSSM}
\par
\begin{figure}[pb]
\begin{center}
\includegraphics[width=0.45\textwidth]{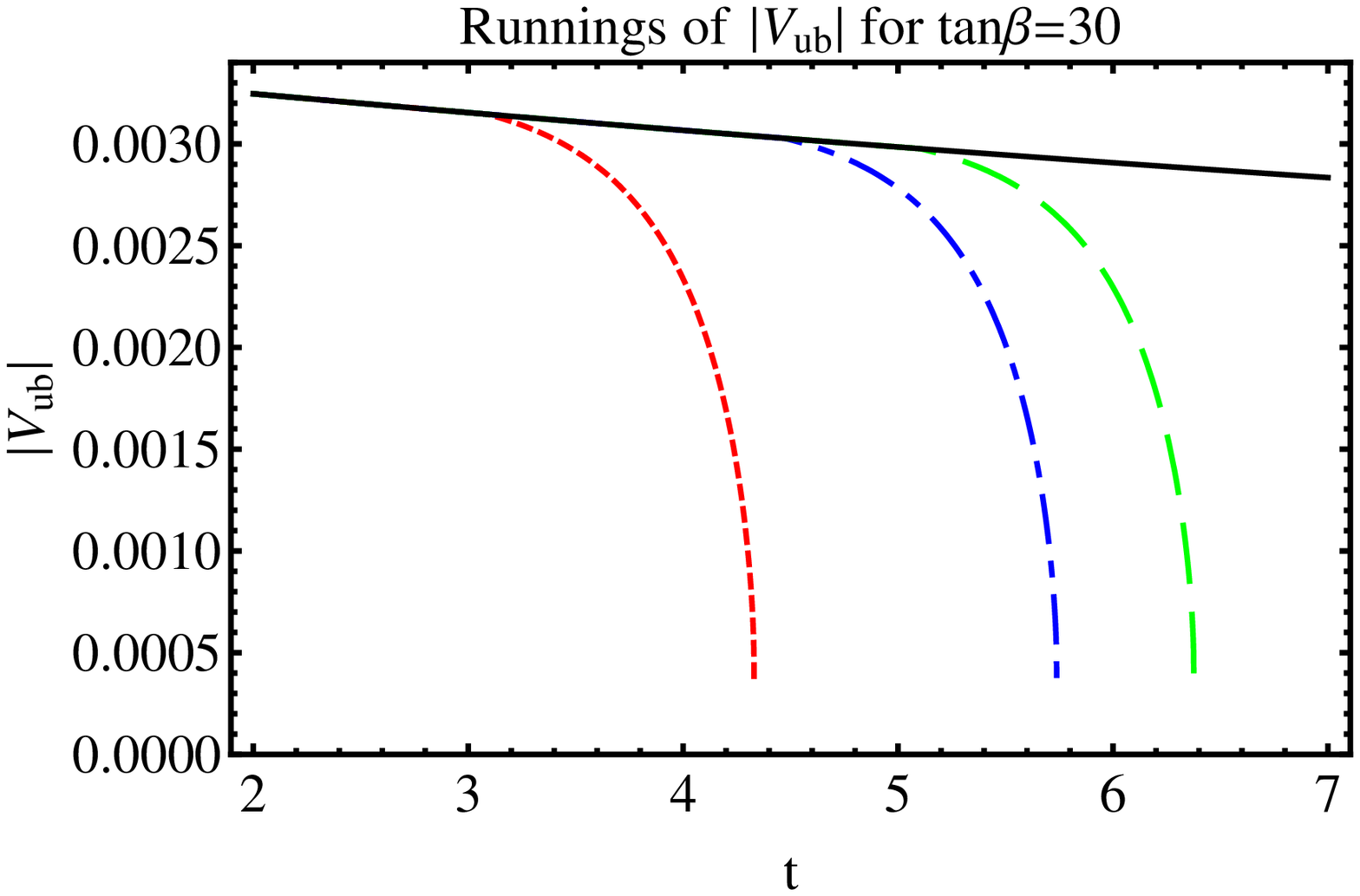}\hspace{0.05\textwidth}
\includegraphics[width=0.45\textwidth]{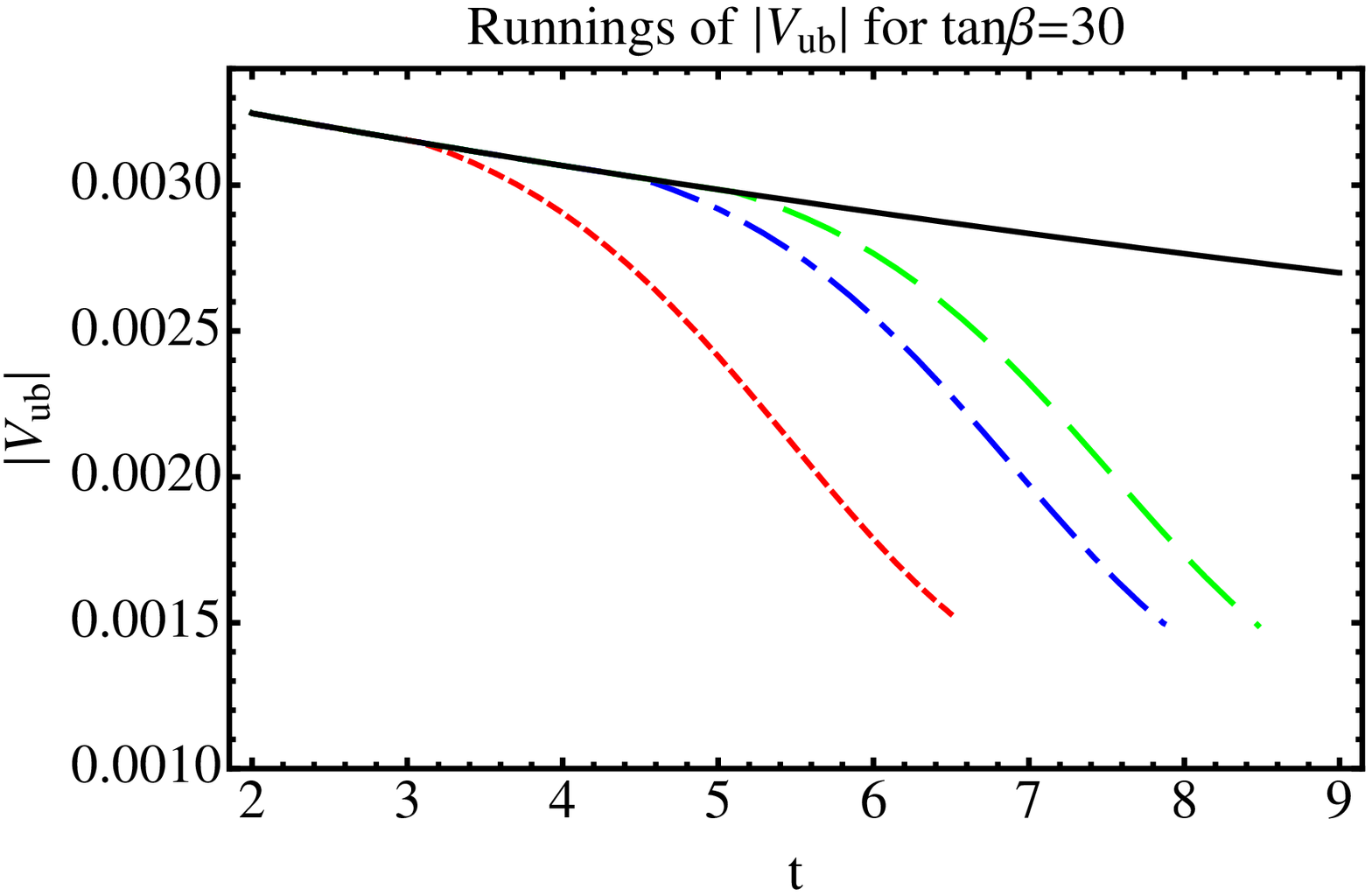}
\vspace*{8pt}
\caption{The CMK matrix elements $|V_{ub}|$ in the 5D MSSM as a function of the scale parameter $t$, for the bulk case (left panel) and brane case (right panel) for $\tan\beta = 30$ where the solid line is the MSSM evolution and for different compactification scales: $R^{-1}$ = 2 TeV (red, dotted line), 8 TeV (blue, dot-dashed line), and 15 TeV (green, dashed line).}
\label{fig:Vub_bulk_brane_30}
\end{center}
\end{figure}

 In Fig.\ref{fig:Vub_bulk_brane_30} we plot the energy dependence of $|V_{ub}|$ from the weak scale all the way up to the high energy scales for different values of compactification radii $R^{-1}$ for the bulk and brane cases in 5D MSSM for $\tan\beta = 30$.

\par The running of the CKM matrix is governed by the terms related to the Yukawa couplings, where $V_{ub} \simeq \theta_{13} e^{-i \delta}$ can be used to observe the mixing angle, $ \theta_{13}$. It decreases with the energy scale in a similar manner regardless of whether $\tan\beta$ is small or large. However, for a large initial value of $f_t$ (small $\tan\beta$), the mixing angles have a more rapid evolution and end in the regime where the top Yukawa diverges and develops a singularity. 

%

\subsection{The Jarlskog parameter}

\subsubsection*{UED SM}

\begin{figure}[pb]
\begin{center}
\includegraphics[width=0.45\textwidth]{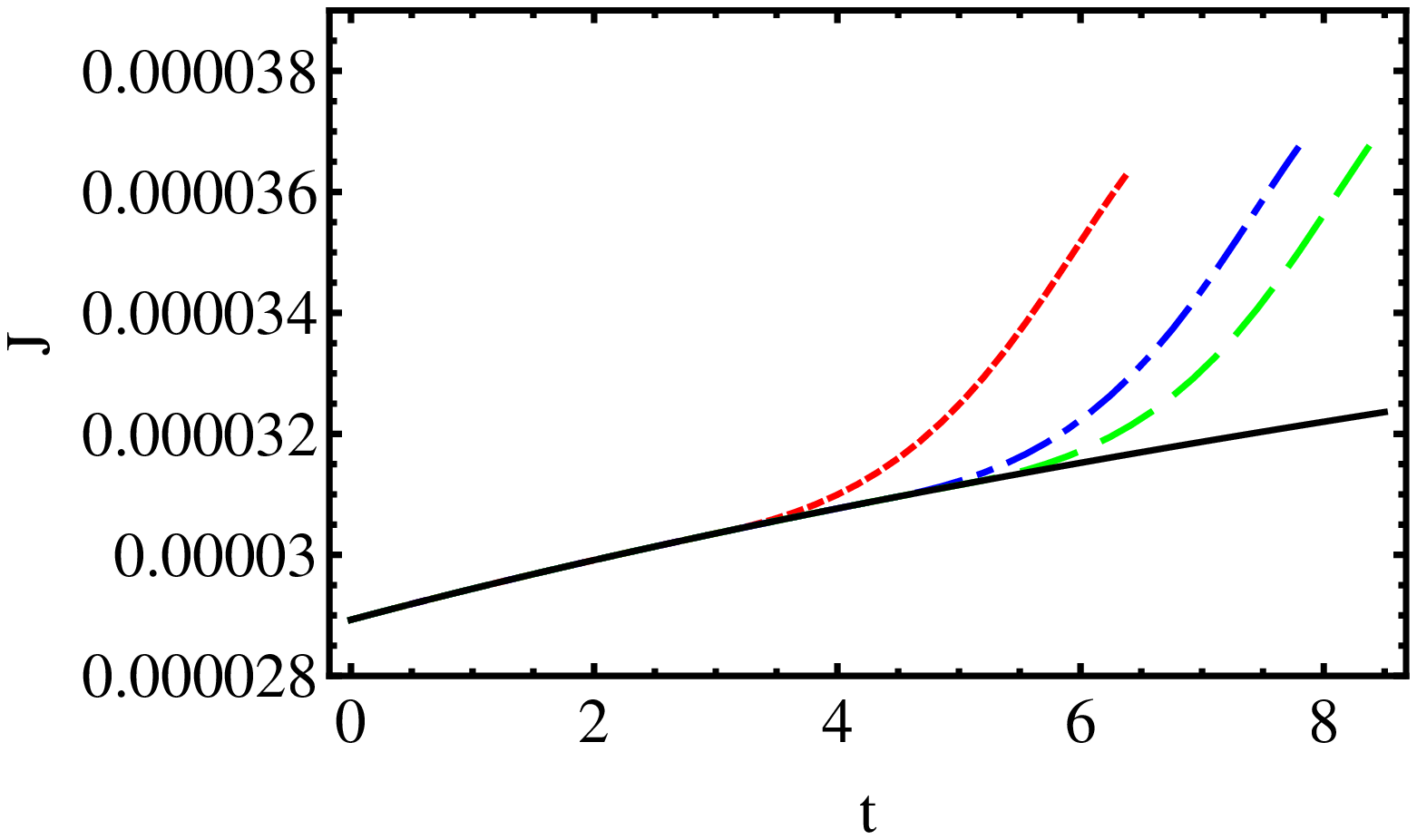}\hspace{0.05\textwidth}
\includegraphics[width=0.45\textwidth]{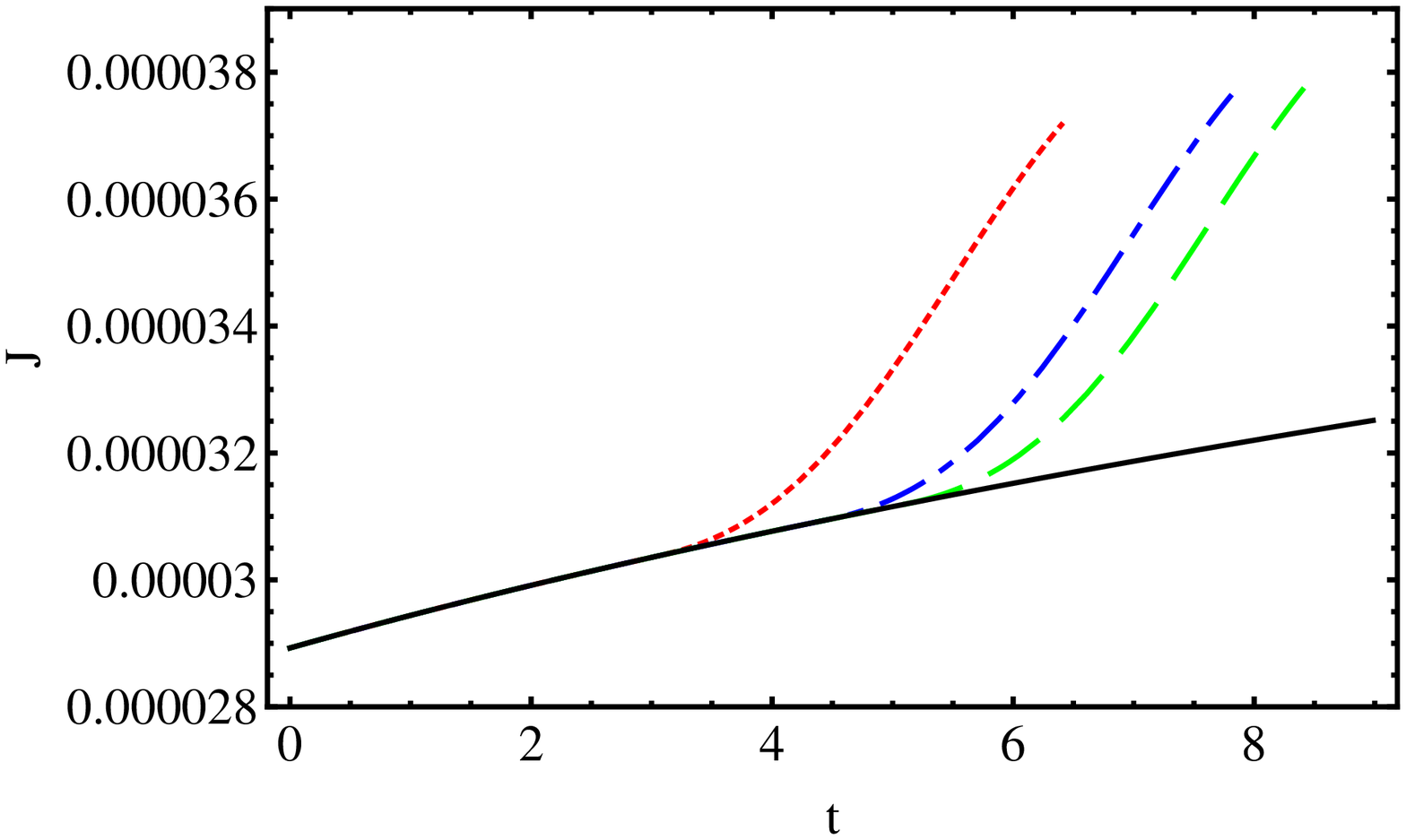}
\vspace*{8pt}
\caption{The Jarlskog parameter $J$ in the UED SM as a function of the scale parameter $t$, for the bulk case (left panel) and the brane case (right panel) where the solid line is the SM, for different compactification scales: $R^{-1}$ = 2 TeV (red, dotted line), 8 TeV (blue, dot-dashed line), and 15 TeV (green, dashed line).}
\label{fig:J_uedsm}
\end{center}
\end{figure}

We next turn our attention to the quark flavor mixing matrix, especially the complex phase of the CKM matrix which characterises CP-violating phenomena. From Fig.\ref{fig:J_uedsm} the variation in the Jarlskog parameter ($J$) becomes very significant. The larger the value of the compactification radius $R$, the faster $J$ evolves to reach its maximum. 

\subsubsection*{5D MSSM}

\par
\begin{figure}[pb]
\begin{center}
\includegraphics[width=0.45\textwidth]{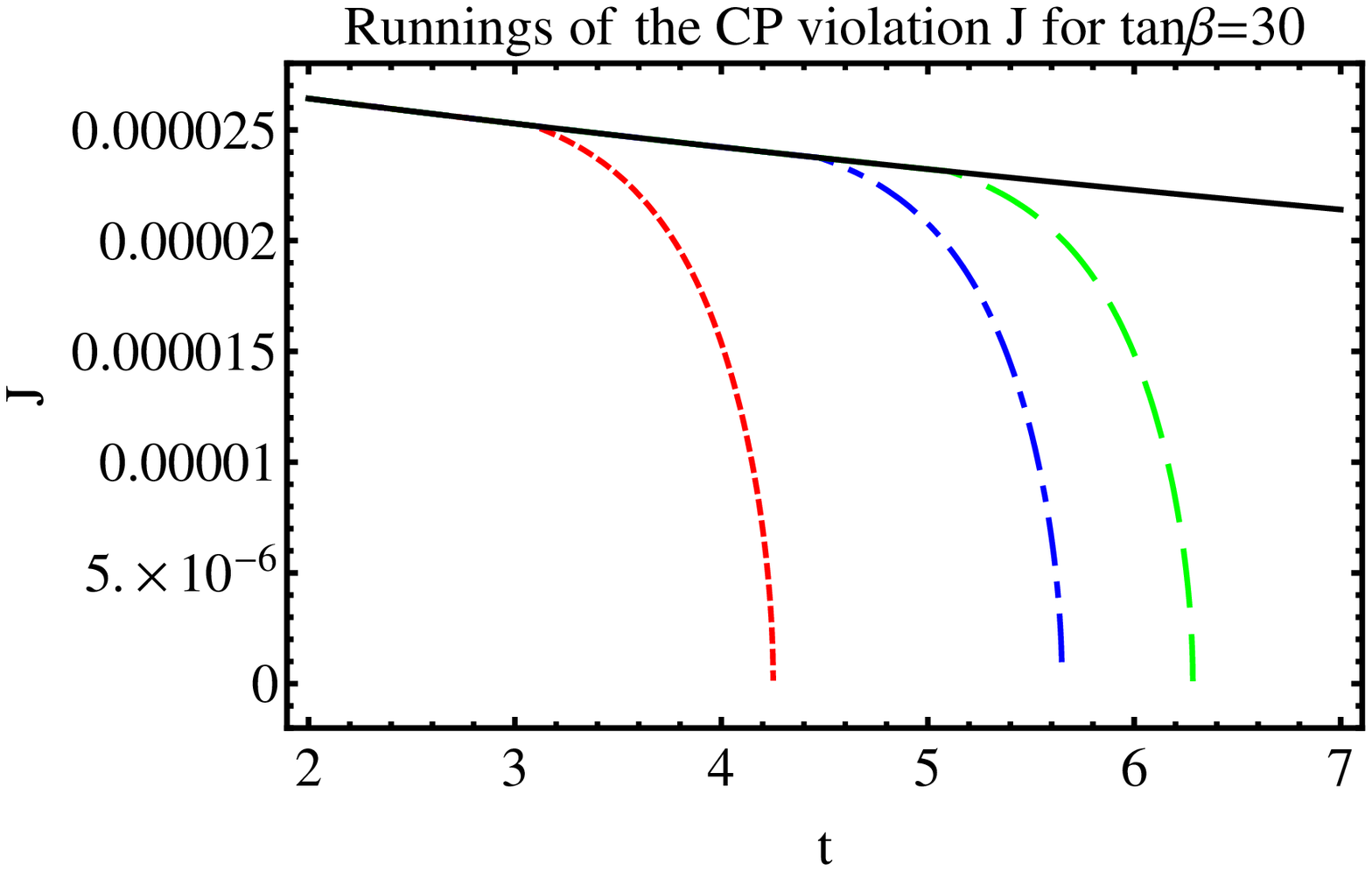}\hspace{0.05\textwidth}
\includegraphics[width=0.45\textwidth]{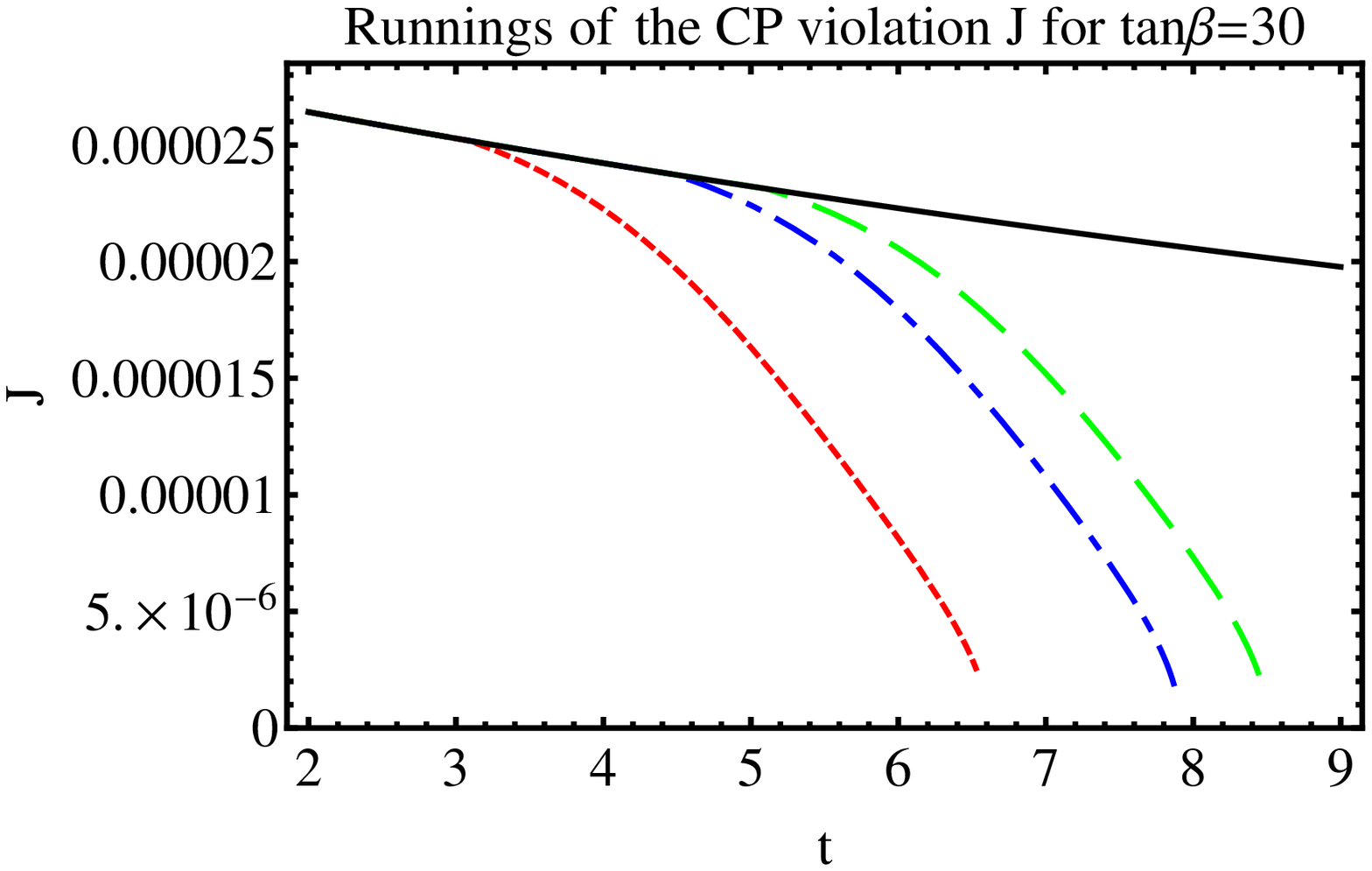}
\vspace*{8pt}
\caption{The Jarlskog parameter $J$ in the 5D MSSM as a function of the scale parameter $t$ for $\tan\beta = 30$, for the bulk case (left panel) and the brane case (right panel) where the solid line is the MSSM evolution and for different compactification scales: $R^{-1}$ = 2 TeV (red, dotted line), 8 TeV (blue, dot-dashed line), and 15 TeV (green, dashed line).}
\label{fig:J_bulk_brane_30}
\end{center}
\end{figure}

 From Fig.\ref{fig:J_bulk_brane_30}, in contrast, the Jarlskog parameter decreases quite rapidly once the initial KK threshold is passed. However, when $\tan\beta$ is large, we have a relatively longer distance between the initial and terminating energy track, the evolution of $J$ can be driven towards zero or even further. 

\par For the matter fields constrained to the brane, in Figs.\ref{fig:Vub_brane} and \ref{fig:J_bulk_brane_30} we observe that the evolutions of these mixing angles and CP violation parameter are decreasing irrespective of whether the top Yukawa coupling grows or not. For small $\tan \beta$ we see similar evolution behaviours for these parameters as in the bulk case. However, as $\tan\beta$ becomes larger, the top Yukawa coupling evolves downward instead of upward. The decreases in these CKM parameters then becomes much milder towards the unification scale; though the reduction to effectively zero in the Jarlskog parameter persists. As a result, for the brane localised matter field scenario, it is more desirable to have a large $\tan\beta$ for theories that are valid up to the gauge coupling unification scale.

%
%

\section{Neutrino parameter evolutions}\label{sec:9}

\par In a similar way to what was done for quark parameters, we can study the evolution of the masses, mixing and phases in the neutrino sector. Indeed the values of the measured mixing angles and the expected sensitivity of future experiments will allow to test at least partially the predicted evolutions of the neutrino parameters.

%

\subsection{Conventions for masses and mixing parameters}

\par The mixing matrix which relates gauge and mass eigenstates is defined to diagonalise the neutrino mass matrix in the basis where the charged lepton mass matrix is diagonal. It is usually parameterised as follows\cite{Maki:1962mu}:
\begin{equation}
U = \left( \begin{array}{ccc}
c_{12} c_{13} & s_{12} c_{13} & s_{13} e^{- i \delta} \\
-s_{12} c_{23} - c_{12} s_{23} s_{13} e^{- i \delta} & c_{12} c_{23} - s_{12} s_{23} s_{13} e^{i \delta} & s_{23} c_{13} \\
s_{12} s_{23} - c_{12} c_{23} s_{13} e^{i \delta} & - c_{12} s_{23} - s_{12} c_{23} s_{13} e^{i \delta} & c_{23} c_{13}
\end{array} \right)
\left( \begin{array}{ccc}
e^{i \phi_1} && \\
& e^{i \phi_2} & \\
&& 1
\end{array} \right) \; , \nonumber
\end{equation}
with $c_{ij} = \cos \theta_{ij}$ and $s_{ij} = \sin \theta_{ij}$ ($ij = 12, 13, 23$).  We follow the conventions of Ref.~\cite{Antusch:2003kp} to extract mixing parameters from the PMNS matrix.

\par Experimental information on neutrino mixing parameters and masses is obtained mainly from oscillation experiments. In general $\Delta m^2_{\mathit{atm}}$ is assigned to a mass squared difference between $\nu_3$ and $\nu_2$, whereas $\Delta m^2_{\mathit{sol}}$ to a mass squared difference between $\nu_2$ and $\nu_1$. The current observational values are summarised in Table~\ref{ta2}. Data indicates that $\Delta m^2_{\mathit{sol}} \ll \Delta m^2_{\mathit{atm}}$, but the masses themselves are not determined. In this work we have adopted the masses of the neutrinos at the $M_Z$ scale as $m_{1} = 0.1$ eV, $m_{2} = 0.100379$ eV, and $m_{3} = 0.11183$ eV, as the {\it normal} hierarchy (whilst any reference to an {\it inverted} hierarchy would refer to $m_{3} = 0.1$ eV, with $m_{3} < m_{1} < m_{2}$ and satisfying the above bounds). For the purpose of illustration, we choose values for the angles and phases as the $M_Z$ scale as: $\theta_{12} = 34^o$, $\theta_{13} = 8.83^o$, $\theta_{23} = 46^o$, $\delta =30^0$, $\phi_1 = 80^o$ and $\phi_2 = 70^o$.

\begin{table}[ph]
\tbl{Present limits on neutrino masses and mixing parameters used in the text. Data is taken from Ref.~\protect{\cite{An:2012eh}} for $\sin^2(2\theta_{13})$, and from Ref.~\protect{\cite{Nakamura:2010zzi}}.}
{\begin{tabular}{@{}cccc@{}} \toprule
Parameter & Value (90\% CL) \\ \hline
$\sin^2(2\theta_{12})$ & $0.861(^{+0.026}_{-0.022})$ \\
$\sin^2(2\theta_{23})$ & $>0.92$ \\
$\sin^2(2\theta_{13})$ & $0.092\pm0.017 $ \\
$\Delta m^2_{\mathit{sol}}$ & $(7.59\pm 0.21)\times 10^{-5}eV^2$ \\
$\Delta m^2_{\mathit{atm}}$ & $(2.43\pm 0.13)\times 10^{-3}$ $eV^2$ \\
\end{tabular}\label{ta2}}
\end{table}

\par The evolution equation for the observables in our 5D MSSM are taken from\cite{ahmad}. As expected $\tan\beta$ plays an important role as all the mixing angles and phases depend on $y_\tau$ (see Appendix C in\cite{ahmad}). However, the new degrees of freedom (the extra-dimensional fields giving rise to KK excitations of the zero modes) become important at energies corresponding to their masses. 

%

\subsection{{$\Delta m^2_{\mathit{sol}}$ and $\Delta m^2_{\mathit{atm}}$}}

\begin{figure}[hpb]
\begin{center}
  \mbox{\epsfxsize=0.5\textwidth\epsffile{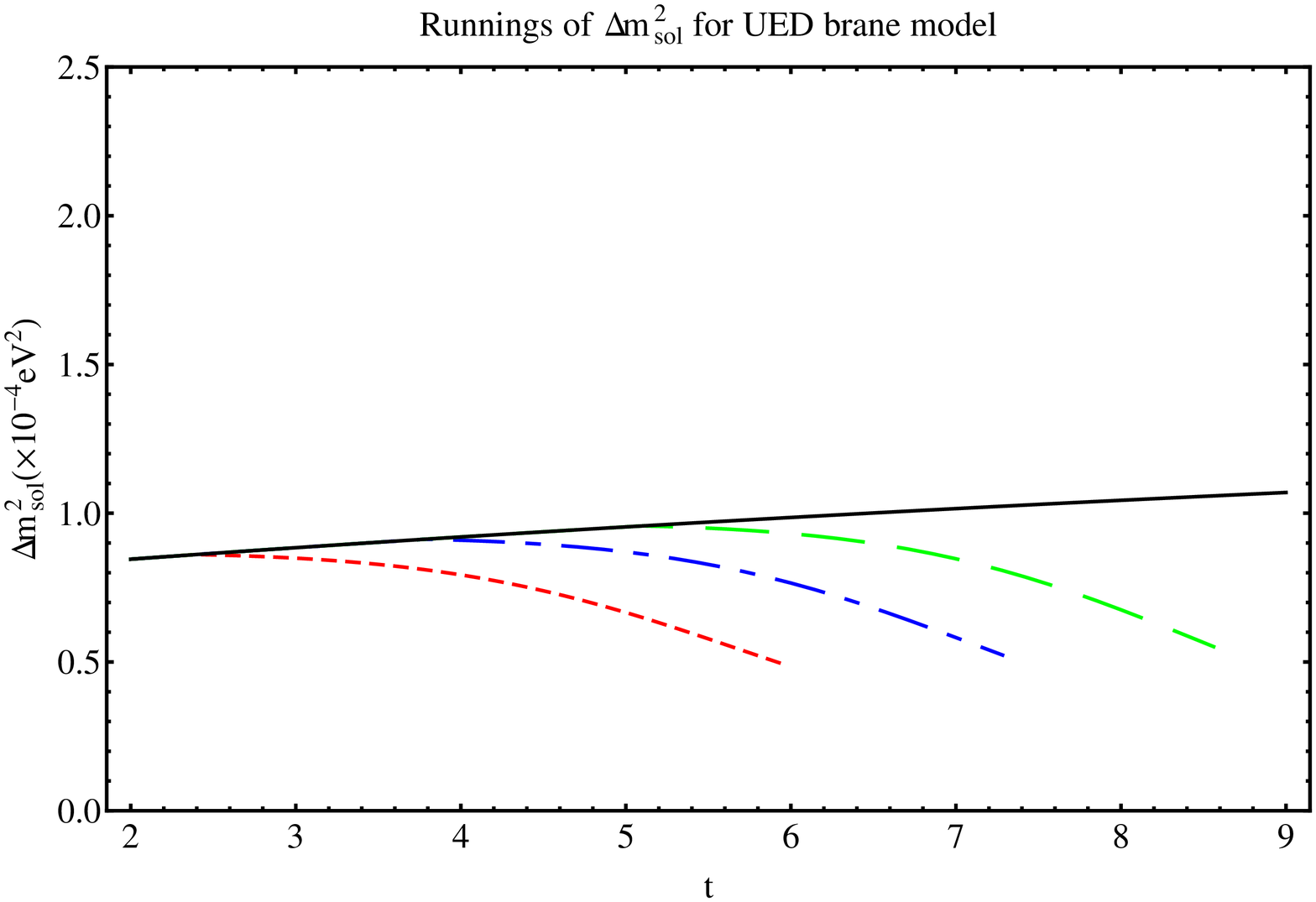} \epsfxsize=0.5\textwidth\epsffile{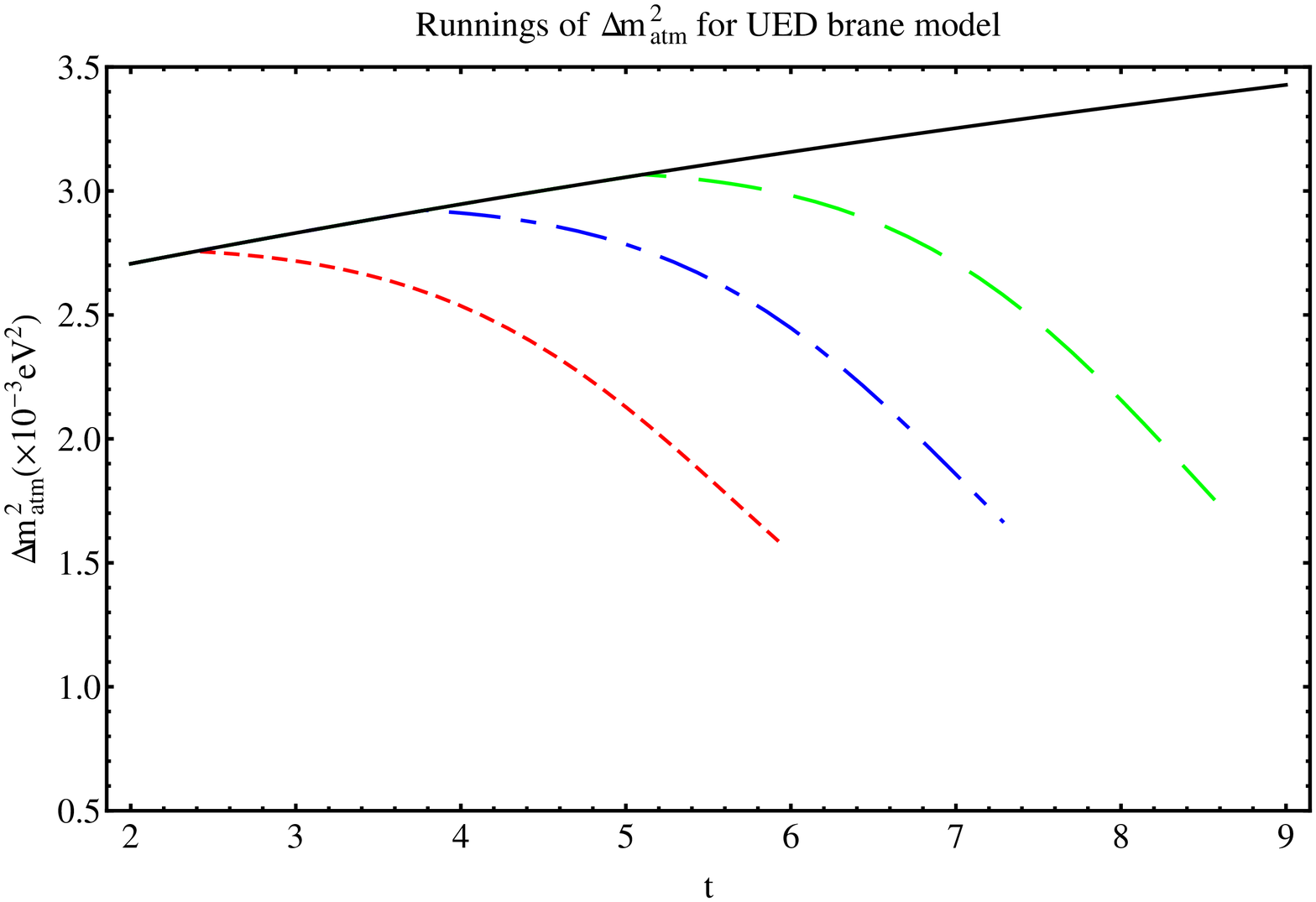}}
  \end{center}
\vspace*{8pt}
\caption{Evolution of $\Delta m_{sol}^2$  (left panel) and  $\Delta m_{atm}^2$ (right panel)  as a function of the scale $t=\ln (\mu/M_Z)$ with matter fields constrained to the brane in the UED SM. The black line is the SM evolution, the red (small dashes) is for $R^{-1}\sim 1$ TeV, the blue (dash-dotted) $R^{-1}\sim 4$ TeV, the green (large dashes) $R^{-1}\sim 15$ TeV.}
\label{ued_brane_deltamass}
\end{figure}
\begin{figure}[hpb]
\begin{center}
  \mbox{\epsfxsize=0.5\textwidth\epsffile{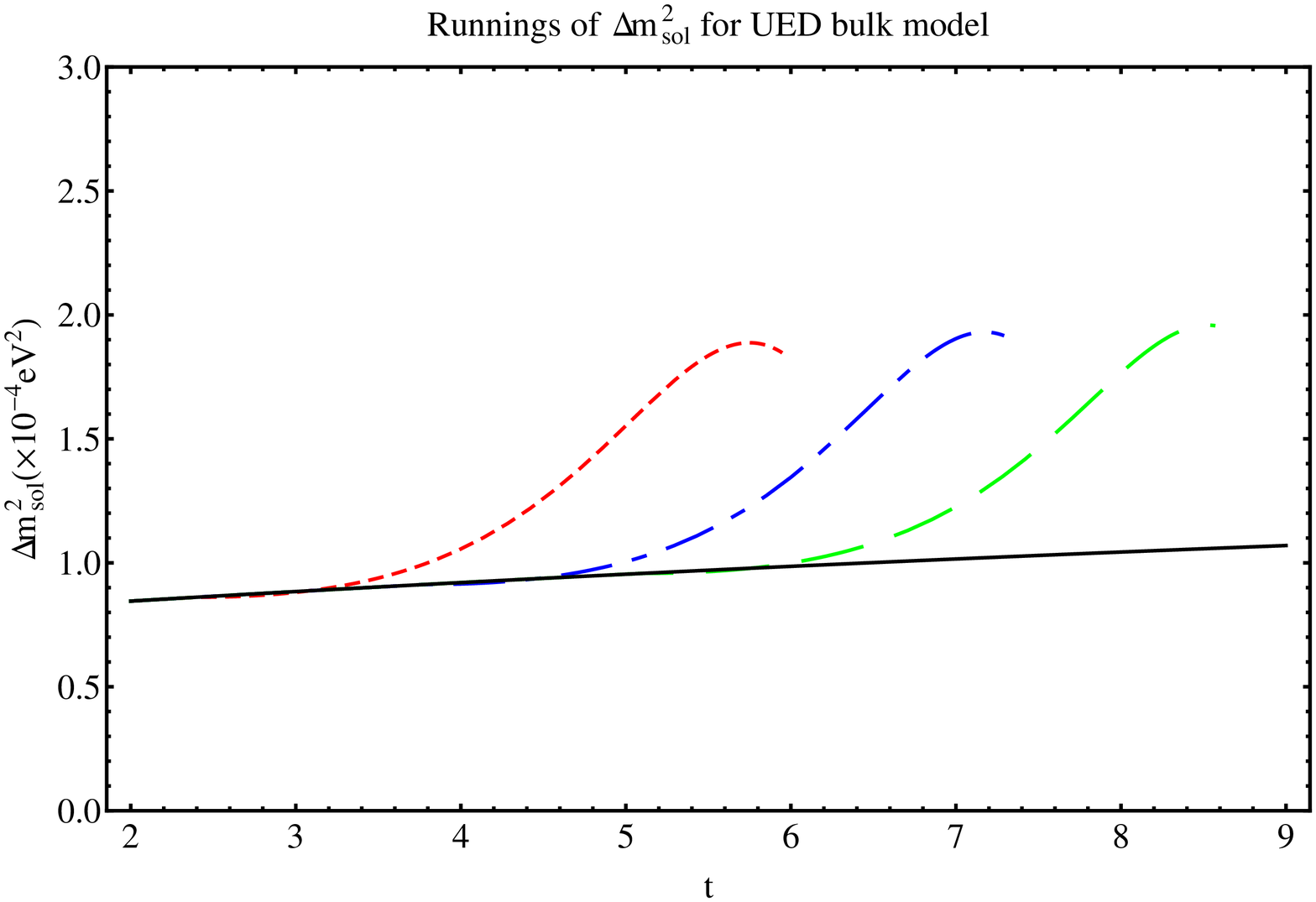} \epsfxsize=0.5\textwidth\epsffile{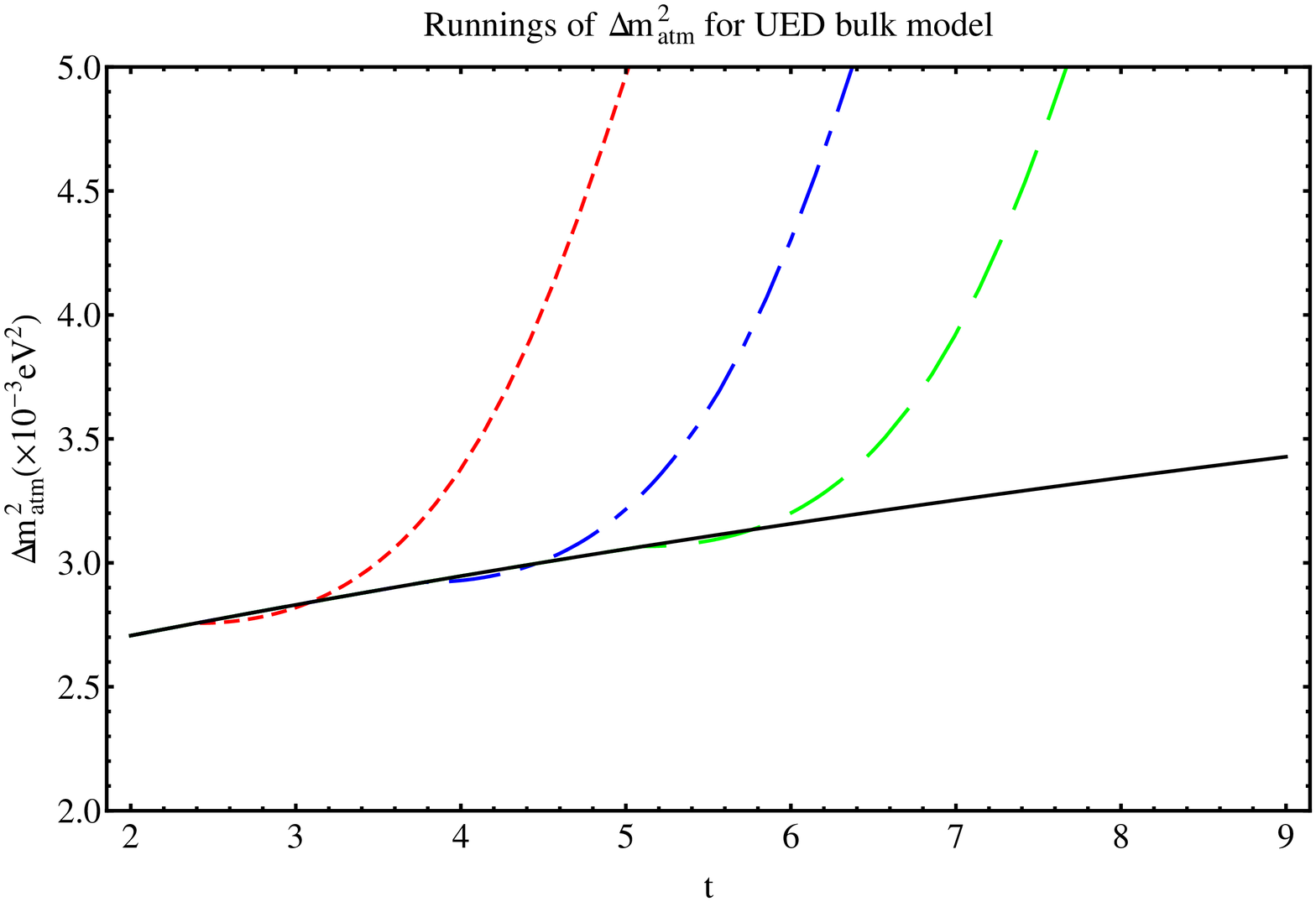}}
  \end{center}
\vspace*{8pt}
\caption{Evolution of $\Delta m_{sol}^2$  (left panel) and  $\Delta m_{atm}^2$ (right panel) as a function of the scale $t=\ln (\mu/M_Z)$ with matter fields in the bulk in the UED SM. The black line is the SM evolution, the red (small dashes) is for $R^{-1}\sim 1$ TeV, the blue (dash-dotted) $R^{-1}\sim 4$ TeV, and the green (large dashes) $R^{-1}\sim 15$ TeV.}
\label{ued_bulk_deltamass}
\end{figure}
\begin{figure}[hpb]
\begin{center}
  \mbox{\epsfxsize=0.5\textwidth\epsffile{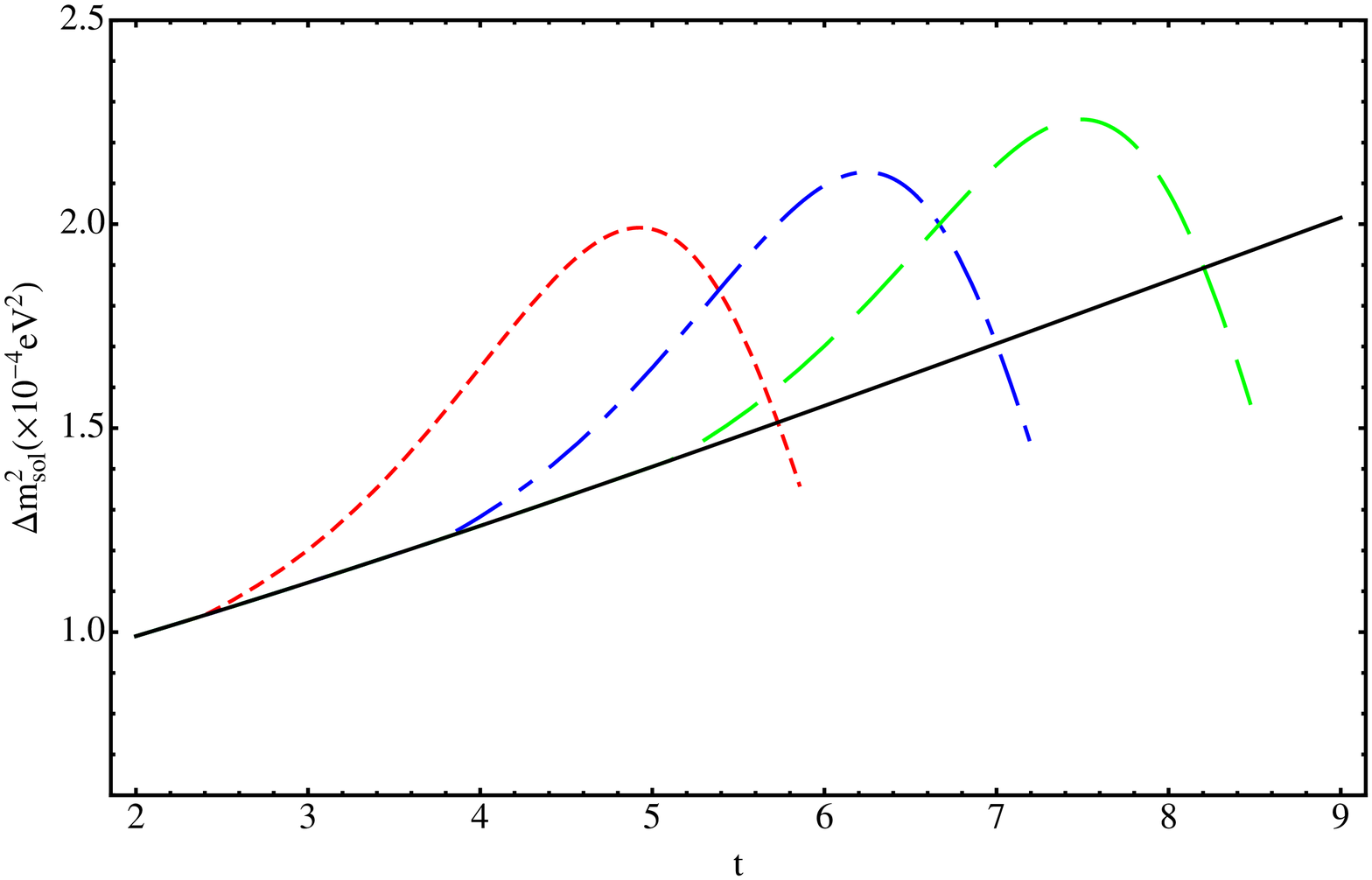} \epsfxsize=0.5\textwidth\epsffile{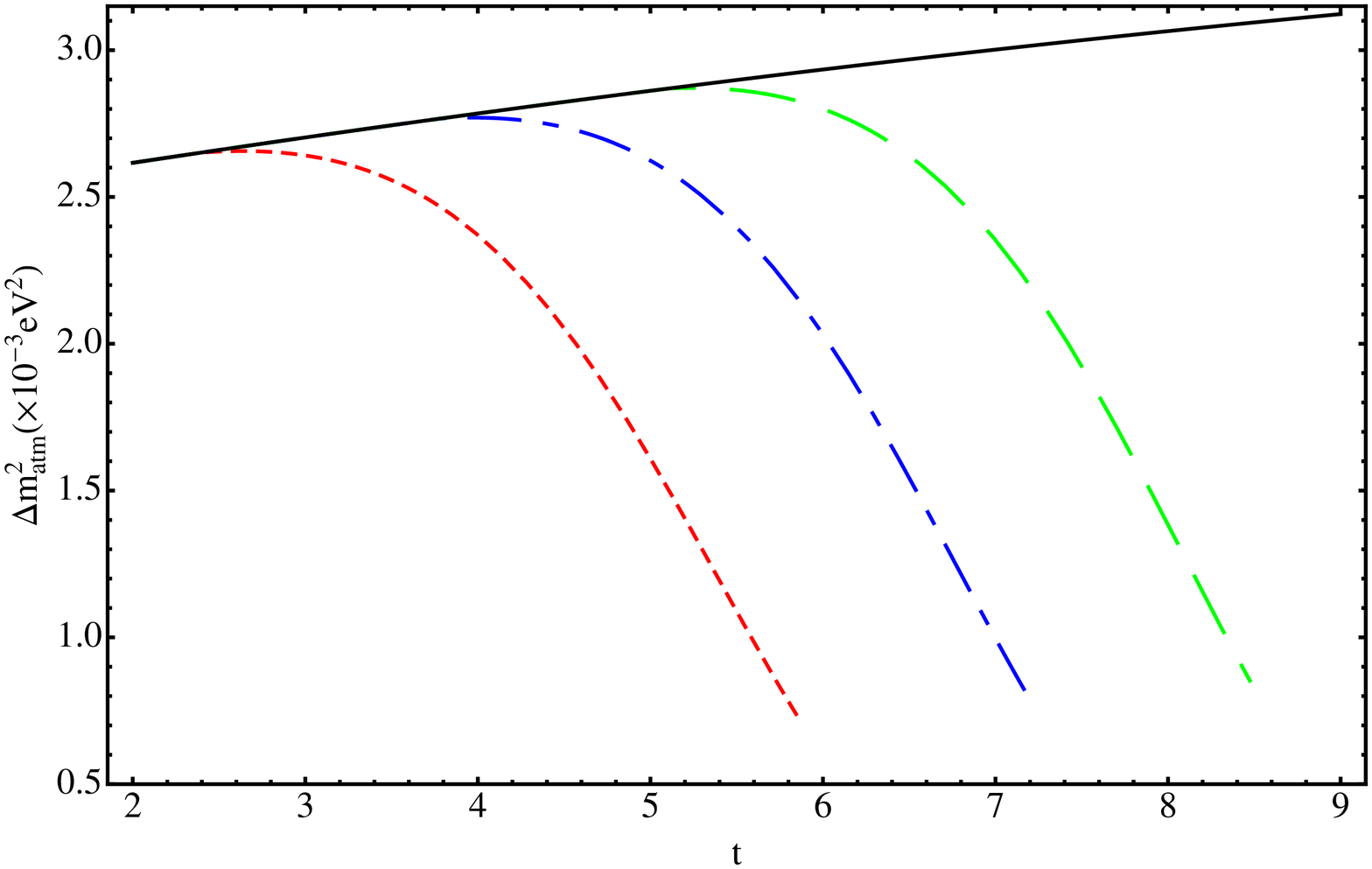}}
  \end{center}
\vspace*{8pt}
\caption{Evolution of $\Delta m_{sol}^2$  (left panel) and  $\Delta m_{atm}^2$ (right panel)  as a function of the scale $t=\ln (\mu/M_Z)$ with matter fields constrained to the brane for $\tan \beta = 30$ in the 5D MSSM. The black line is the MSSM evolution, the red (small dashes) is for $R^{-1}\sim 1$ TeV, the blue (dash-dotted) $R^{-1}\sim 4$ TeV, the green (large dashes) $R^{-1}\sim 15$ TeV.}
\label{brane_deltamass}
\end{figure}
\begin{figure}[hpb]
\begin{center}
  \mbox{\epsfxsize=0.5\textwidth\epsffile{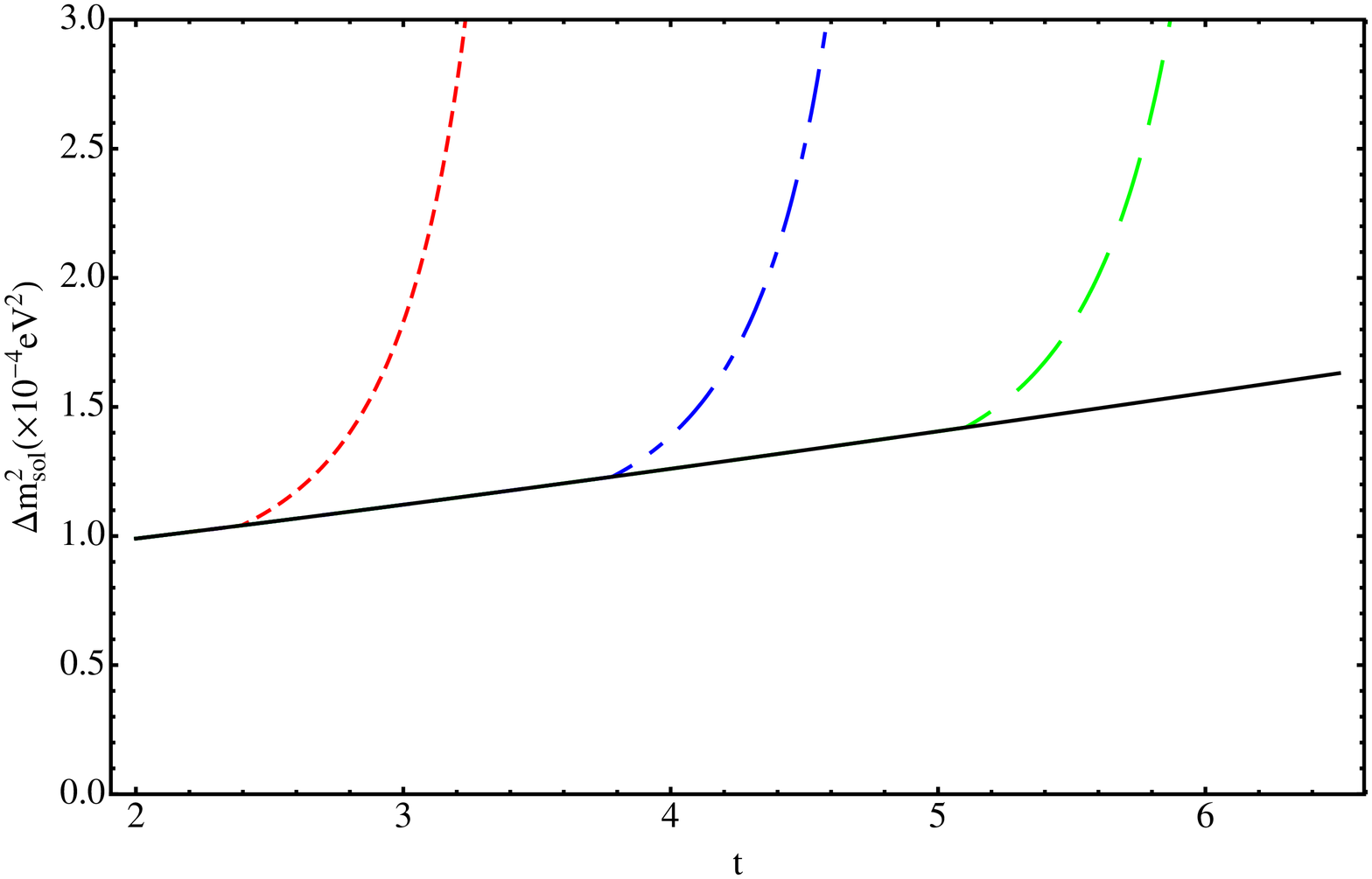} \epsfxsize=0.5\textwidth\epsffile{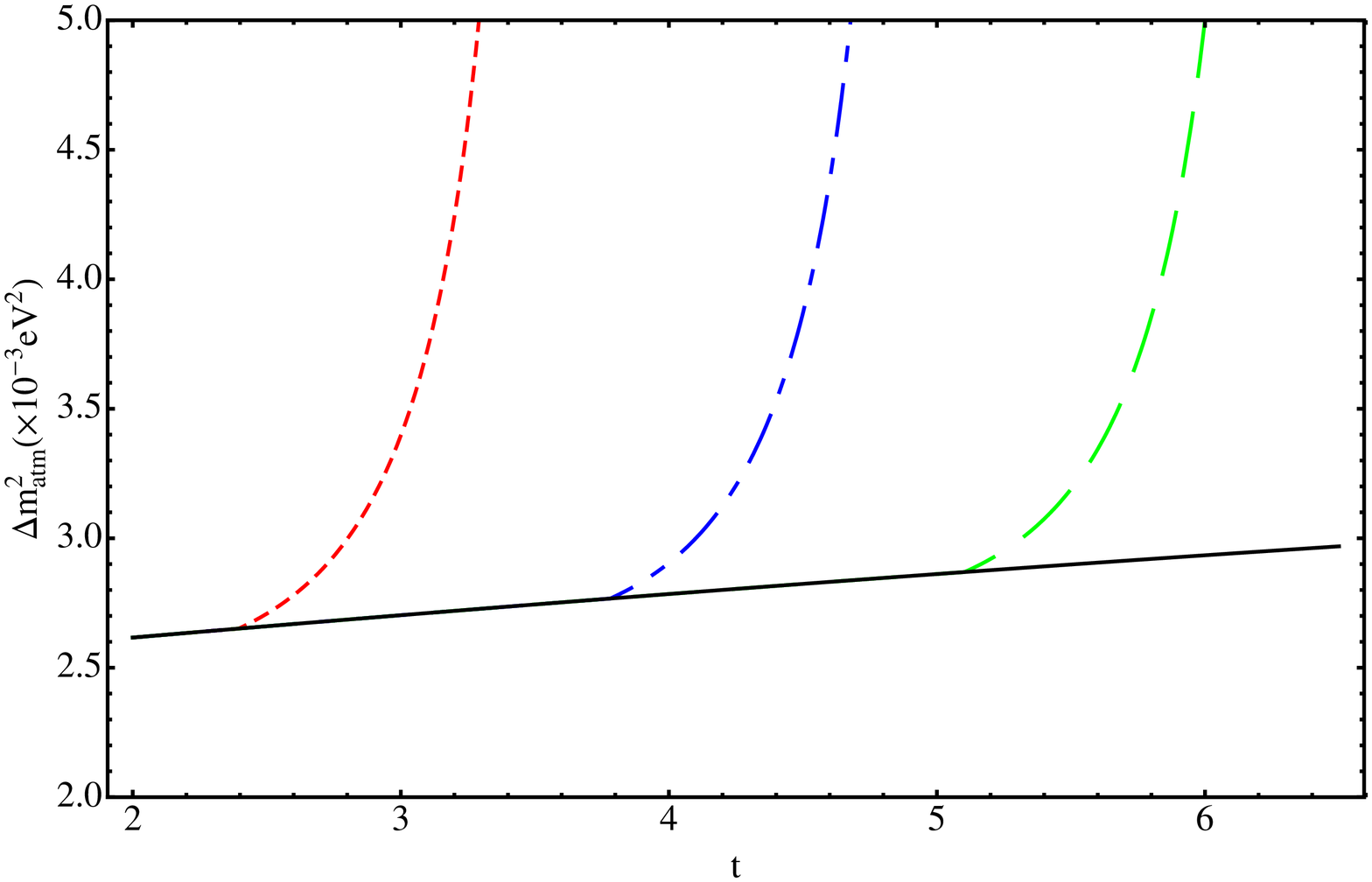}}
  \end{center}
\vspace*{8pt}
\caption{Evolution of $\Delta m_{sol}^2$  (left panel) and  $\Delta m_{atm}^2$ (right panel) as a function of the scale $t=\ln (\mu/M_Z)$ with matter fields in the bulk for $\tan \beta = 30$ in the 5D MSSM. The black line is the MSSM evolution, the red (small dashes) is for $R^{-1}\sim 1$ TeV, the blue (dash-dotted) $R^{-1}\sim 4$ TeV, and the green (large dashes) $R^{-1}\sim 15$ TeV. The evolution is towards a non-perturbative regime, where the Yukawa coupling develops a Landau pole and the effective theory becomes invalid.}
\label{bulk_deltamass}
\end{figure}

\par For the UED SM, we see different behaviour for the brane case Fig.\ref{ued_brane_deltamass} and bulk case Fig. \ref{ued_bulk_deltamass}. Once the KK threshold is reached, both $\Delta m_{sol}^2$ and $\Delta m_{atm}^2$ decrease with increasing energy in the brane case, but they increase with the energy in the bulk case for the different radii of compactification. The evolution of masses depends on the evolution of $y_{\tau}$ and $k$ coupling and the RG runnings in the UED SM bulk model are generally larger than those in UED SM brane model. This is due to the fact that the coefficient $C_1=0$ in the brane model (sec.(\ref{UEDBranecoef})) and $2(S(t)-1)$ in the bulk model (sec.(\ref{UEDBulkcoef})) and also there is difference in $\alpha$ in the two equations due to the trace of charged-fermion Yukawa couplings in bulk model whereas such a contribution does not exist in brane model due to the absence of fermion KK excitations (see the $T$ term in sec.(\ref{SMcoef})). This lead to the increasing of observables in the bulk case and the decreasing in the brane case.

\par For the 5D MSSM, in general, in the brane case, the evolution has the same form for the three masses $m_1$, $m_2$, $m_3$. This leads to a reduction of up to a factor of two for the masses at $t=6$ (for a large radius, $R^{-1}=1$ TeV) with respect to the MSSM values at low energies.

\par The situation is more involved when analysing the mass squared differences. We plot in Figs.\ref{brane_deltamass} and \ref{bulk_deltamass} the evolution of $\Delta m_{sol}^2$ and $\Delta m_{atm}^2$ both for the matter fields on the brane and for all fields in the bulk for tan$\beta$=30 and different radii of compactification. In the brane case different behaviours as a function of the energy scale are possible as a relatively large interval in energy range is allowed for the effective theory. As explicitly illustrated in Fig. \ref{brane_deltamass}, the relevant radiative corrections controlled by the gauge fields in secs.(\ref{MSSMcoef}, \ref{5Dbranecoef}) become dominant as energy goes up, which tends to reduce mass splitting, and an approximately degenerate neutrino masses spectrum at the high energy scale $m_1 \approx m_2 \approx m_3$ becomes favourable. This is in contrast with the MSSM, where the neutrino mass splitting becomes large at an ultraviolet cut-off. Therefore, it is very appealing that the neutrino mass splitting at low energy could be attributed to radiative corrections resulting from a degenerate pattern at a high energy scale. In Fig.\ref{bulk_deltamass}, the bulk case tends to a non-perturbative regime, where the unitarity bounds of the effective theory are reached much faster and only a much shorter running can be followed using the effective theory. 

\subsection{Mixing angles}
\par
\begin{figure}[hpb]
\begin{center}
  \mbox{\epsfxsize=0.5\textwidth\epsffile{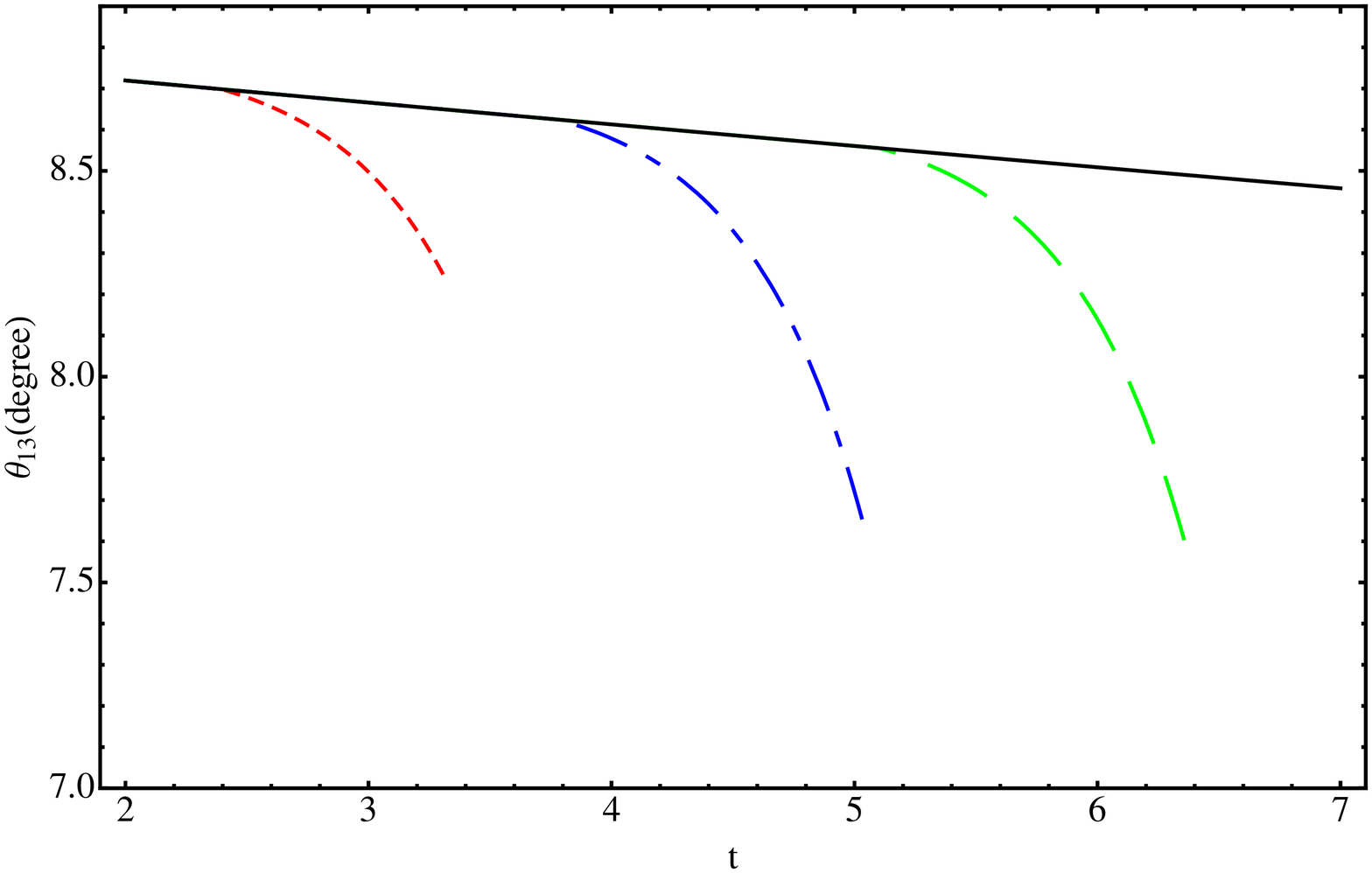} \epsfxsize=0.5\textwidth\epsffile{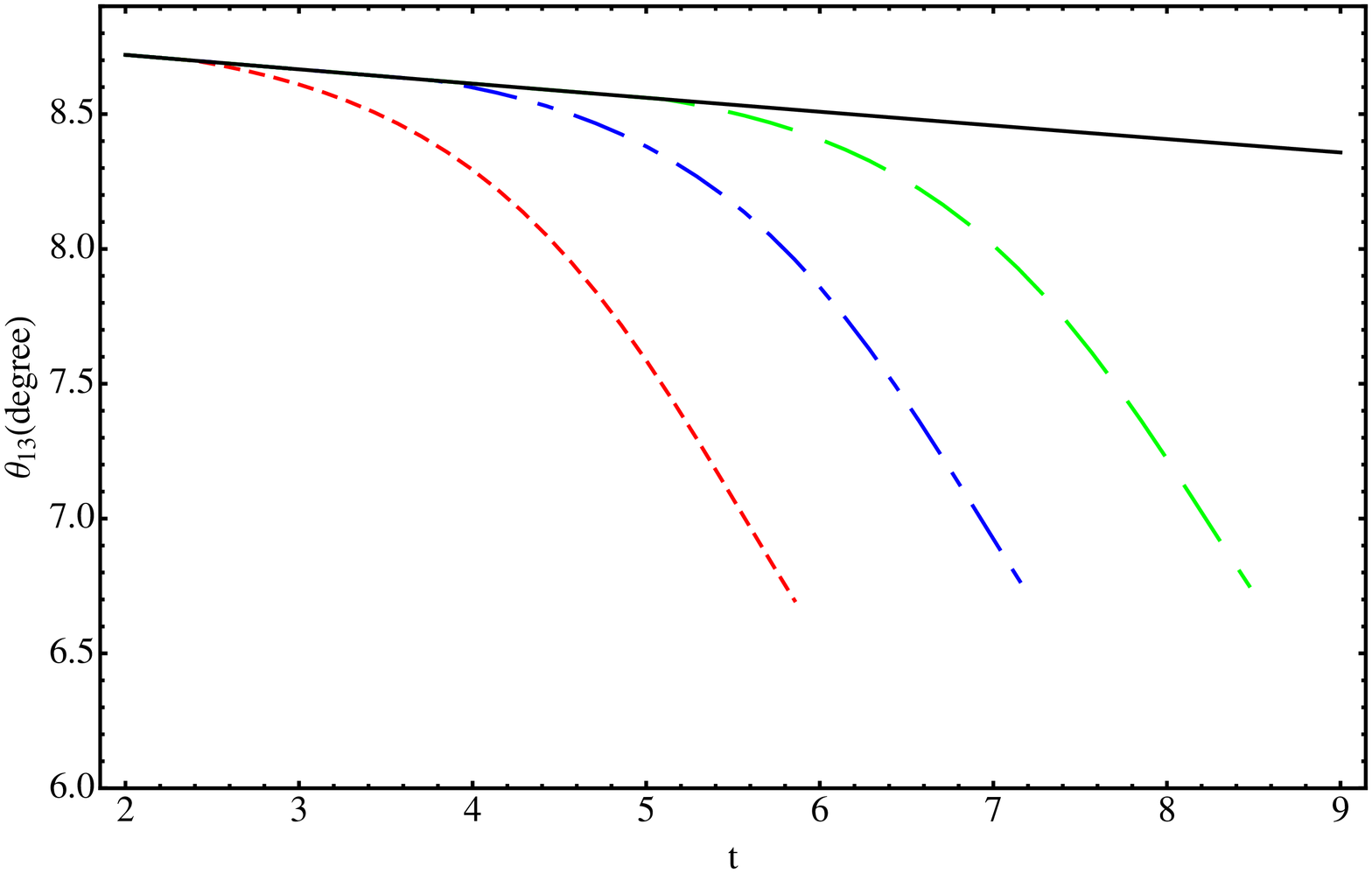}}
  \end{center}
\vspace*{8pt}
\caption{Evolution of $\theta_{13}$  as a function of the parameter $t=\ln (\mu /M_Z)$ for $\tan \beta = 30$ with matter fields in the bulk (left panel) and constrained to the brane (right  panel) in the 5D MSSM. The black line is the MSSM evolution, the red (small dashes) is for $R^{-1}\sim 1$ TeV, the blue (dash-dotted) $R^{-1}\sim 4$ TeV, and the green (large dashes) $R^{-1}\sim 15$ TeV.}
\label{theta13_bulk}
\end{figure}

\begin{figure}[hpb]
\begin{center}
  \mbox{\epsfxsize=0.5\textwidth\epsffile{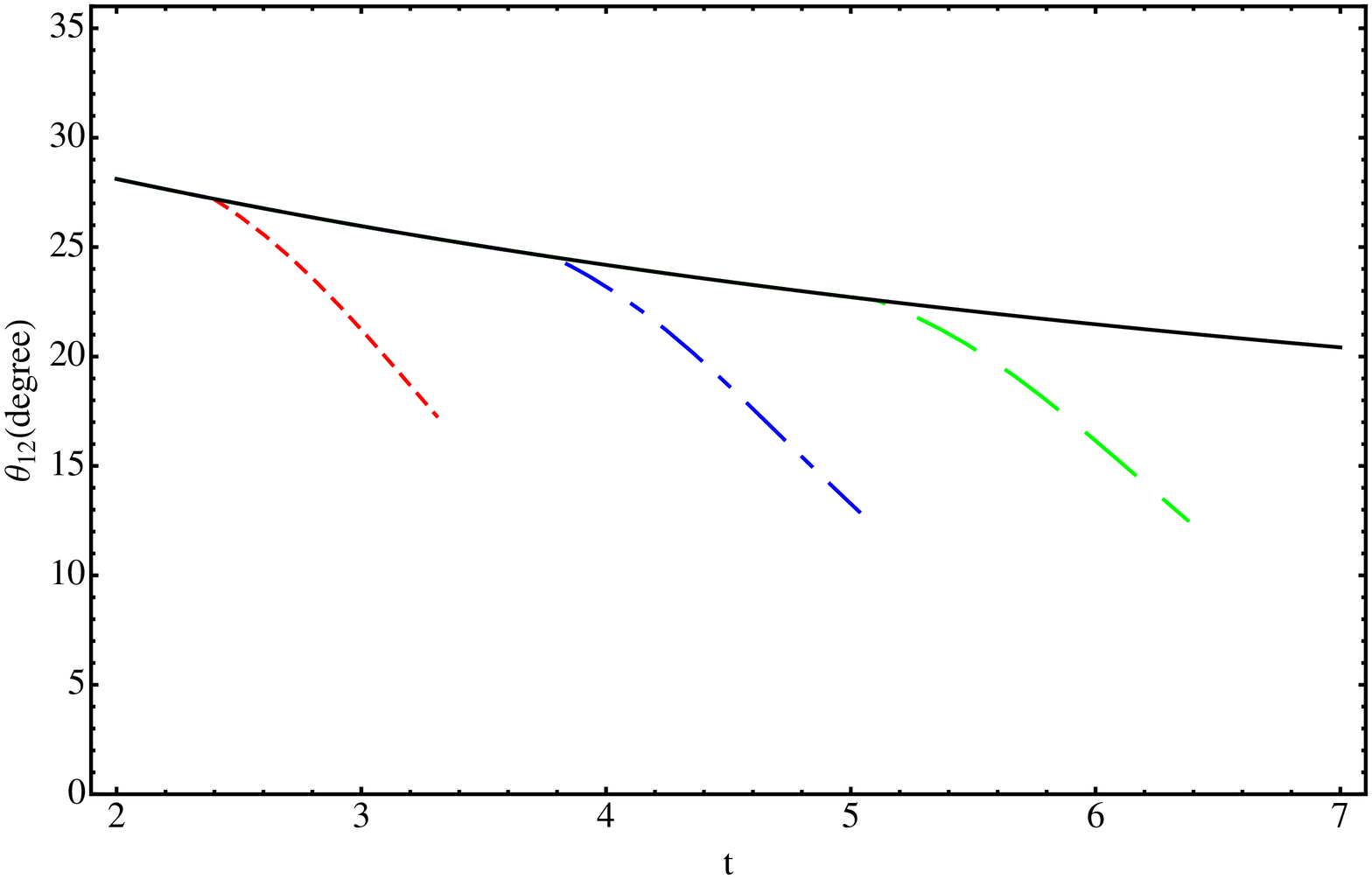} \epsfxsize=0.5\textwidth\epsffile{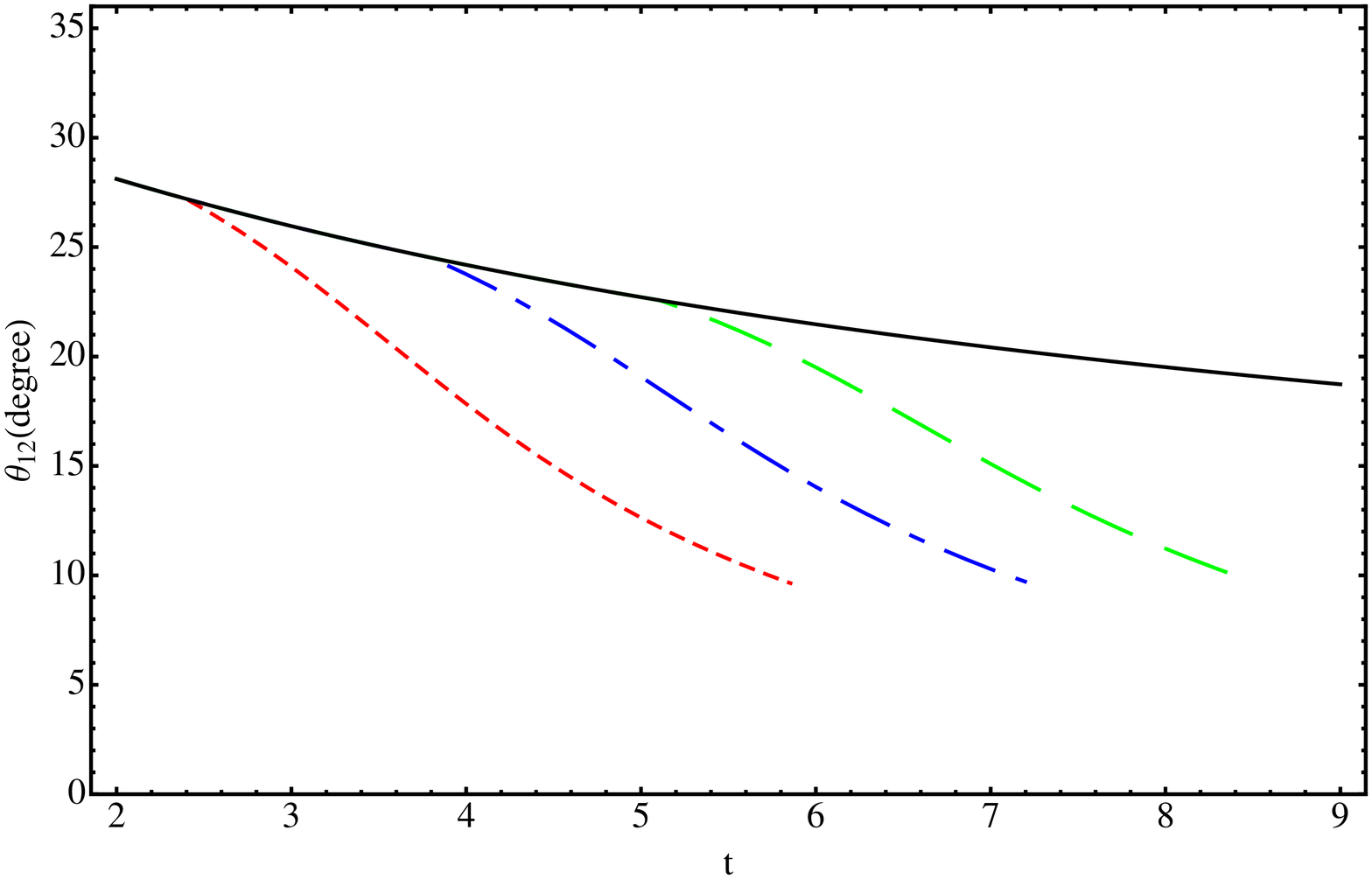}}
  \end{center}
\vspace*{8pt}
\caption{Evolution of $\theta_{12}$  in the bulk (left panel) and on the brane (right panel) as a function of the scale $t=\ln (\mu/M_Z)$ for $\tan \beta = 30$ in the 5D MSSM. The black line is the MSSM evolution, the red one (small dashes) is for $R^{-1}\sim 1$ TeV, the blue (dash-dotted) $R^{-1}\sim 4$ TeV, and the green (large dashes) $R^{-1}\sim 15$ TeV.}
\label{theta12_tanb30}
\end{figure}

 Concerning the evolution of the mixing angles, as can be seen in Figs.\ref{theta13_bulk}--\ref{theta12_tanb30}, in the UED bulk and brane cases, we have very small variation from the SM case because there is no dependence on tan$\beta$ and there is no quadratic term of $S(t)$ in the RGEs. However the mixing angles variation is more significant in the 5D MSSM in which the largest effect is for $\theta_{12}$. 

As observed, due to the large quadratic term of $S(t)$ in the beta function, the $\theta_{12}$ has a rapid and steep variation in the bulk case. However, for the brane case, it has a relatively longer evolution track with the $\theta_{12}$ then being pulled further down until the termination point (where the effective theory becomes invalid). In contrast, the running of $\theta_{13}$ and $\theta_{23}$ is much milder. However, a running to $\theta_{13} = 0$ cannot be observed in any situation.

\subsection{$\delta$ phase}

\par The running of the Dirac phase $\delta$ in the UED SM case is very small.
The variation is stable and similar for the bulk and brane cases, there is very small deviation from the SM because all other mixing angles vary only by small quantities and the coefficient $C$ which appear in the variation of $\delta$ (see Appendix D of \cite{ahmad}) are linear in $S(t)$ and there is no dependence on $\tan \beta$.
\begin{figure}[hpb]
\begin{center}
  \mbox{\epsfxsize=0.5\textwidth\epsffile{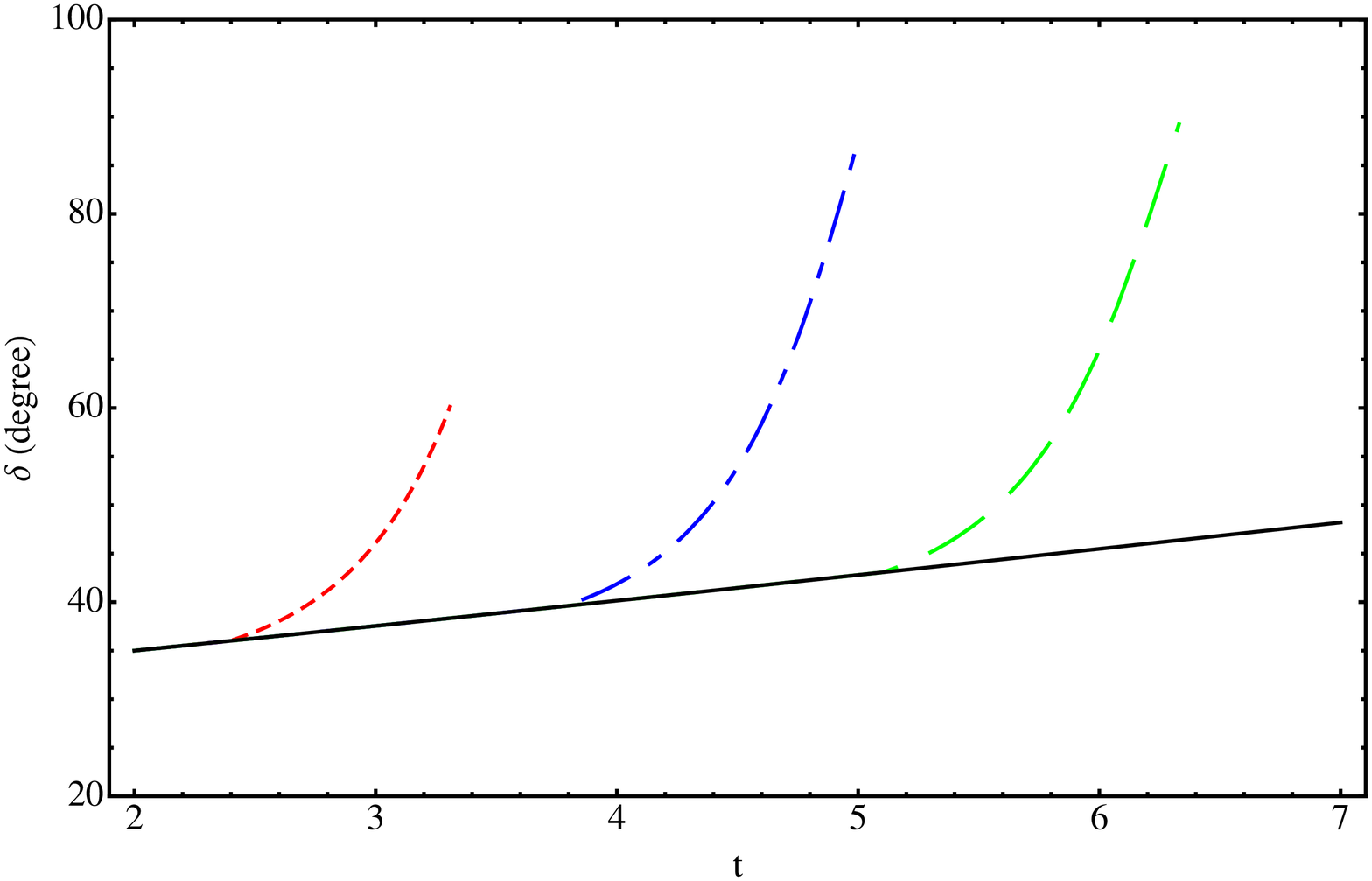} \epsfxsize=0.5\textwidth\epsffile{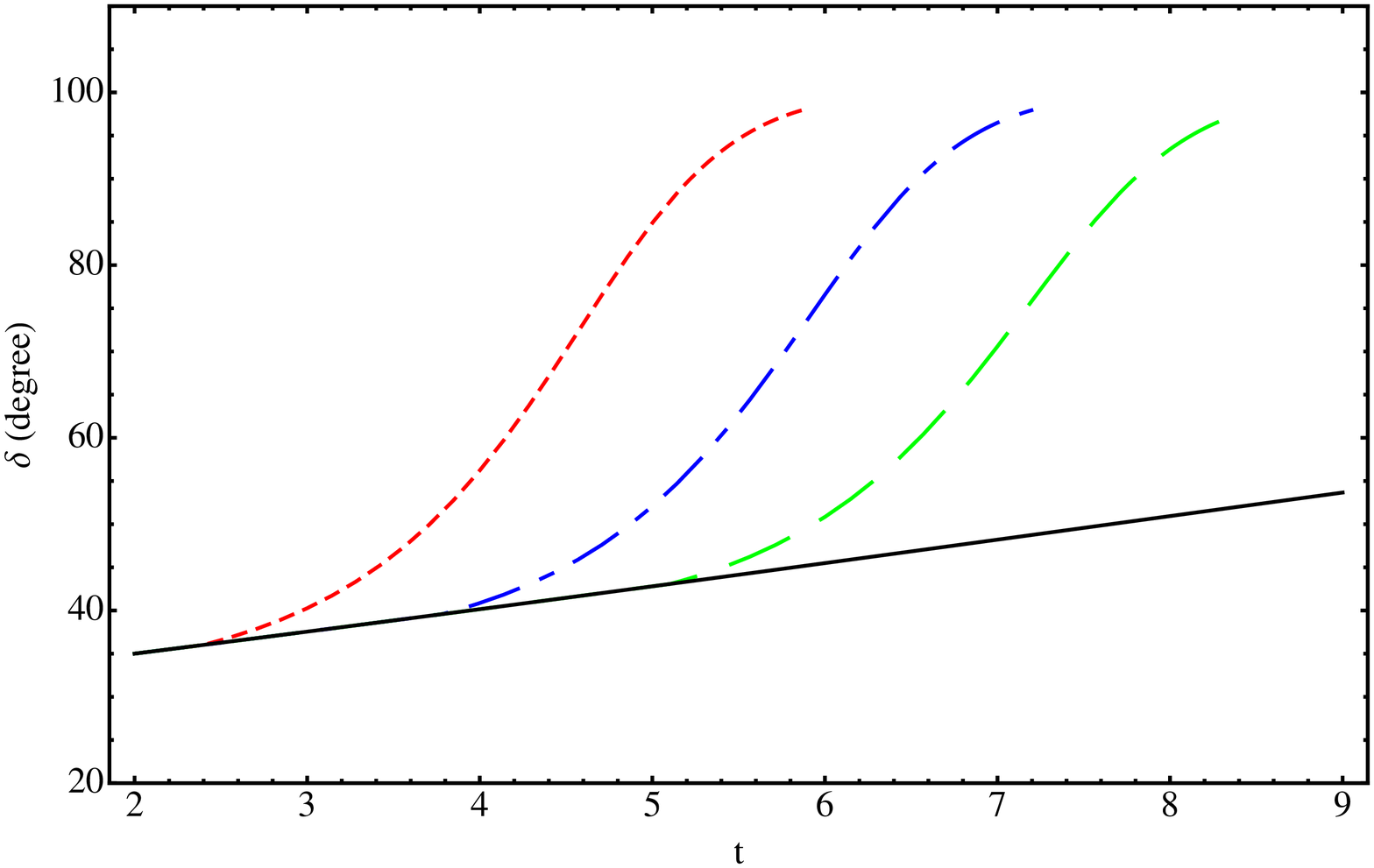}}
  \end{center}
\vspace*{8pt}
\caption{Evolution of the phase $\delta$ as a function of the parameter $t=\ln (\mu/M_Z)$ for $\tan \beta = 50$ with matter fields in the bulk (left panel) and constrained to the brane (right  panel) in the 5D MSSM. The black line is the MSSM evolution, the red one (small dashes) is for $R^{-1}\sim 1$ TeV, the blue (dash-dotted) $R^{-1}\sim 4$ TeV, and the green (large dashes) $R^{-1}\sim 15$ TeV.}
\label{delta_bulk_brane_50}
\end{figure}
\par Noting that the Dirac phase $\delta$ determines the strength of CP violation in neutrino oscillations. In the 5D MSSM, the runnings we include follow the general features presented in Fig.\ref{delta_bulk_brane_50}, with large increases possible once the first KK threshold is crossed.

\par The recent results from the Daya bay and RENO reactor experiments have established non zero values of $\theta_{13}$. 
Therefore, the leptonic CP violation characterised by the Jarlskog invariant
$J \sim \sin {\theta _{12}}\cos {\theta _{12}}\sin {\theta _{23}}\cos {\theta _{23}}\sin {\theta _{13}}{\cos ^2}{\theta _{13}}\sin \delta$
becomes promising to be measured in the future long baseline neutrino oscillation experiments. For leptogenesis related to the matter-antimatter asymmetry, we should note that the parameters entering the leptogenesis
mechanism cannot be completely expressed in terms of low-energy neutrino mass parameters. Note that in some specific models the parameters of the PMNS matrix (which contains CP asymmetry effects) can be used \cite{Raidal:2008jk, Buchmuller:2003gz}. Here, the CP-violating effects induced by the  renormalisation group corrections could lead to values of the CP asymmetries large enough for a successful leptogenesis, and the models predicting maximum leptonic CP violation, or where the CP-violating phase $\delta$ is not strongly suppressed, become especially appealing. Specific models with large extra dimensions in which leptogenesis is relevant at low scale can also be found in Ref. \cite{Gu:2010ye}.

\par The running of the mixing angles are entangled with the CP-violating phases\cite{ahmad}. The phases $\phi_1$ and $\phi_2$ do not affect directly the running of the masses, while the phase $\delta$ has a direct effect on the size of ${dm}/{dt}$, although its importance is somewhat reduced by the magnitude of $\theta_{13}$. For further discussions of the correlation between these phases and mixing angles, refer to Refs \cite{Antusch:2003kp, Luo:2012ce} for details.

\par Finally, whilst the above results and analysis were for the normal hierarchy of neutrino masses, we did also review the inverted hierarchy, where from an analysis of the equations presented in the Appendices of \cite{ahmad} we obtain the same features and results for neutrino mass runnings (though with different initial values at the $M_Z$ scale). As such, the figures for $\Delta m_{sol}^2$ and $\Delta m_{atm}^2$ remain unchanged. Possible changes in the angles and phases arise from the different signs for the $(m_j - m_i)/(m_j + m_i)$ terms present in each evolution equation, where the $\theta_{12}$ results remain approximately the same, and the small runnings of $\theta_{13}$ and $\theta_{23}$ would be up rather than down.

%
%

\section{Summary and Outlook}\label{sec:10}

\par The present review of the renormalisation group evolution of the masses, mixing angles and phases of the UED models in the quark and lepton sectors brings together the results obtained in the recent years in this subject using a common notation. The important physical points are discussed and the equations are written in compact way to show the unified approach to the different sectors of these models. For more technical details we refer to the existing literature.

\par We plot their running up to the gauge unification scale only when relevant, since the introduction of new ultraviolet cutoff becomes imperative due to the scalar potential stability condition, and beyond this scale new physics should appear. In contrast, in the UED brane model, the physics parameters have a full running till the gauge unification scale, since the Higgs self coupling evolution has a finite value which thus excludes the vacuum stability concern and validates the theory up its full scale \cite{Liu:2012me}.

\par The UED model has substantial effects on the hierarchy between the quark and lepton sectors and provides a very desirable scenario for grand unification. The scale deviation of renormalisation curves from the usual SM one depends closely on the value of the compactified radius R. The smaller the radius is, the higher the energy scale we need to differentiate the UED curve from the SM one. A comparison between theoretical predictions and experimental measurements will be available once the LHC will be running at its full centre of mass energy. This will set limits on the parameters of the UED model, and a precise determination of $J$, $|V_{ub}|$ or $|V_{cb}|$ at high energy may lead to a discrimination between the SM and extra dimensional models.

\par In the case of the 5D MSSM, we have reviewed the behaviour of the evolution equations for the quark and neutrino sector in a minimal supersymmetric model with one extra-dimension. For quarks, the 5D MSSM scenarios with matter fields in the bulk or on the brane, give both results with small or no quark flavor mixings at high energies, especially for the mixings with the heavy generation. The evolution of these CKM parameters have a rapid variation prior to reaching a cut-off scale where the top Yukawa coupling develops a singularity point and the model breaks down. For the brane localised matter fields model, we can only observe similar behaviour for small values of $\tan\beta$, while for large $\tan\beta$, the initial top Yukawa coupling becomes smaller, the gauge couplings then play a dominant role during the evolution of the Yukawa couplings, and therefore the Yukawa couplings decrease instead of increasing. Concerning the neutrino sector, the evolution equations for the mixing angles, phases, $\Delta m_{sol}^2$ and $\Delta m_{atm}^2$, within the two distinct scenarios, is also considered. A larger $\tan \beta$ typically leads to larger renormalisation group corrections. Neutrino masses evolve differently in the two models due to the sign of the (different) dominant contributions in the bulk and in the brane cases. 

\par To conclude this review we would like to note some of the remaining incomplete areas of investigation in the study of quark and lepton sector runnings in UED models. Whilst we have reviewed the simplest SM and MSSM UED models, other alternative extra-dimensional geometries exist. Note that our two scenarios of all matter fields freely propagating in the bulk or brane localised represent the only possibilities for calculating unitary CKM or PMNS matrices, where extensions to the runnings of Yukawas with different numbers of matter fields in the bulk or brane are trivial extensions of the equations already reviewed here. Alternative extra-dimensional geometries are still to be investigated, such as 2UED models (preliminary work \cite{Ohlsson:2012hi} contains errors and are incomplete studies of these sectors) or situations with warped Randall-Sundrum style extra-dimensions; though warp factors provide an additional problem of vertex factors now depending on the KK numbers, and so the equations would become of a completely different form to the ones provided here (excepting extreme limiting cases).


%
%

\section*{Acknowledgements}
ASC would like to thank AD, AT and the IPNL for their hospitality during his stay in Lyon, where the first stage of this work was
performed.

%
%

\end{document}